\documentclass
[aps,pra,reprint,showpacs,floatfix,amsmath,amssymb,superscriptaddress,longbibliography,footinbib]
{revtex4-2}

\usepackage{graphicx}
\usepackage{dcolumn}
\usepackage{bm}
\usepackage[colorlinks=true,urlcolor=blue,citecolor=blue,linkcolor=blue]{hyperref}
\usepackage{braket}
\usepackage{comment}
\usepackage{mathtools}
\usepackage{multirow}
\usepackage{lineno}
\usepackage{siunitx}
\usepackage{orcidlink}
\usepackage{color}
\usepackage{footmisc}
\graphicspath{{./Fig/}}
\newcommand{\g}{{}^1S_0}
\newcommand{\e}{{}^3P_2}
\newcommand{\clock}{{}^1S_0 \leftrightarrow {}^3P_2}

\newcommand{\dn}[2]{\delta \nu_{#1}^{#2}}

\newcommand{\w}[1]{w^{#1}}
\newcommand{\dr}[1]{\delta \langle r^2 \rangle ^{#1}}

\newcommand{\ddr}[1]{\left[\delta \langle r^2 \rangle ^2 \right]^{#1}}

\newcommand{\Dn}[1]{\bm{\delta \nu_{#1}}}

\newcommand{\Dr}[0]{\bm{\delta \langle r^2 \rangle}}
\newcommand{\Drr}[0]{\bm{\delta \langle r^4 \rangle}}
\newcommand{\Ddr}[0]{\bm{\left[\delta \langle r^2 \rangle ^2 \right]}}

\newcommand{\DA}[0]{\bm{\delta A}}

\begin{document}

\title{Precision Isotope-Shift Measurement of the $\clock$ Transition \\and a Multi-Transition King-Plot Consistency Test}

\author{Taiki Ishiyama\,\orcidlink{0000-0002-5699-3713}}
\thanks{These authors contributed equally to this work. Corresponding author: ishiyama.taiki.v81@kyoto-u.jp}
\affiliation{Department of Physics, Graduate School of Science, Kyoto University, Kyoto 606-8502, Japan}
\author{Koki Ono\,\orcidlink{0000-0002-1145-0424}}
\thanks{These authors contributed equally to this work. Corresponding author: ishiyama.taiki.v81@kyoto-u.jp}
\affiliation{Department of Physics, Graduate School of Science, Kyoto University, Kyoto 606-8502, Japan}
\affiliation{HIKARI-COOL Kyoto, Kyoto 606-8502, Japan}
\author{Reiji Asano\,\orcidlink{0009-0005-3861-6235}}
\affiliation{Department of Physics, Graduate School of Science, Kyoto University, Kyoto 606-8502, Japan}
\author{Hokuto Kawase\,\orcidlink{0009-0000-5075-9048}}
\affiliation{Department of Physics, Graduate School of Science, Kyoto University, Kyoto 606-8502, Japan}
\author{Tetsushi Takano\,\orcidlink{0000-0002-0406-4605}}
\affiliation{Department of Physics, Graduate School of Science, Kyoto University, Kyoto 606-8502, Japan}
\affiliation{The Hakubi Center for Advanced Research, Kyoto University, Kyoto 606-8502, Japan}
\author{Ayaki Sunaga\,\orcidlink{0000-0002-3802-680X}}
\thanks{Present address: Department of Chemistry, Faculty of Science, Hokkaido University, Sapporo, Hokkaido 060-0810, Japan.}
\affiliation{ELTE, Eötvös Loránd University, Institute of Chemistry, Pázmány Péter sétány Budapest 1/A 1117, Hungary}
\author{Yasuhiro Yamamoto\,\orcidlink{0000-0003-1478-6043}}
\affiliation{Accelerator Laboratory, High Energy Accelerator Research Organization (KEK), Tsukuba 305-0801, Japan}
\author{Minoru Tanaka,\orcidlink{0000-0001-8190-2863}}
\affiliation{Department of Physics, Graduate School of Science, The University of Osaka, Toyonaka 560-0043, Japan} 
\author{Yoshiro Takahashi\,\orcidlink{0000-0001-7607-7387}}
\affiliation{Department of Physics, Graduate School of Science, Kyoto University, Kyoto 606-8502, Japan}
\date{\today}


\begin{abstract}
We present precision spectroscopy and isotope shift measurement of the $\clock$ clock transition in neutral ytterbium (Yb) atoms.
By revealing a magic wavelength at $905.4(2)$ nm, we successfully achieve the atomic spectrum narrower than 100~Hz.
The interleaved clock operation between isotopes allows us to determine isotope shifts of four bosonic isotope pairs at Hz-level uncertainties, which is combined with those of other four ultra-narrow transitions in Yb and Yb$^+$ to construct the King plot.
We show that a rich dataset spanning multiple transitions enables a test of a restricted isotope-shift ansatz comprising the leading mass and field shifts, one effective factorized higher-order Standard Model contribution, and a new-particle contribution.
Within this ansatz, the five-transition data are inconsistent with the new-particle contribution serving as the sole additional contribution over the mass ranges $m_\phi \leq 6.5$~keV and 18~keV $\leq m_\phi <$ 100 keV, while the previous works rely on another bound set by the neutron scattering and $(g-2)_e$ measurements.
This work paves the way for the effective use of precision isotope shift data in the King plot analysis and stimulates further measurements in Yb and other elements.
\end{abstract}
\maketitle


\paragraph*{Introduction.}
In spite of the great success of the Standard Model (SM) in particle physics, it is also recognized as incomplete, since some phenomena, such as dark matter and dark energy, are still insufficiently explained~\cite{Navas2025}.
Various new-physics models beyond the SM have been proposed and tested using a wide variety of experimental techniques, ranging from high-energy accelerators to low-energy precision measurements~\cite{Safronova2018-xl}.
Among the latter, the King plot approach based on isotope shift (IS) measurements has attracted much attention as a sensitive probe to a new Yukawa-type interaction between electrons and neutrons~\cite{Frugiuele2017-mp, Berengut2018-su, Berengut2025-yy}.
If a nonlinearity is observed in a two-dimensional (2D) King plot~\cite{King1963-sb} or its generalization to higher dimensions~\cite{Mikami2017-oh}, this might be caused by the new particle.
On the other hand, a higher-order SM effect can also violate these linearities~\cite{Mikami2017-oh, Munro-Laylim2022-ho}; thus, a careful analysis is indispensable to get a constraint on the new particle.

Ytterbium (Yb) is one of the most actively studied elements for the King plot approach~\cite{Counts2020-manual, Ono2022-oy, Hur2022-manual, Figueroa2022-cm, Door2025-manual, Ishiyama2026-wd}, as it has many narrow and ultra-narrow optical transitions as well as five nuclear-spin-less bosonic stable isotopes.
Notably, two transitions in Yb ${}^1S_0 \leftrightarrow {}^3P_0$ (578 nm)~\cite{Ono2022-oy} and $4f^{13}5d6s^2 \ (J=2)$ (431 nm)~\cite{Ishiyama2026-wd}, as well as two in Yb$^+$ ${}^2S_{1/2} \leftrightarrow {}^2D_{5/2}$ (411 nm) and ${}^2F_{7/2}$ (467 nm)~\cite{Door2025-manual} have been measured with uncertainties of 3, 8, 5, and 16~Hz, respectively.
The 3D generalized King plot for the transition set of (411, 431, 578) shows a significant nonlinearity of $85 \sigma$~\cite{Ishiyama2026-wd}.
Assuming that the new particle is only one source of this huge nonlinearity, the favored region is obtained in a mass and coupling-constant plane of the new particle.
However, the existence of the new particle, signaled by the finite value of the coupling strengths, is denied by the argument that this region conflicts with the already-existing upper bound from other laboratory-based experiments, specifically the neutron scattering~\cite{Leeb1992-vi, Pokotilovski2006-ct, Nesvizhevsky2008-mf} and $(g-2)_e$ measurements~\cite{Fan2023-cz}.
Since there is an argument about the limitation on applicability of this bound~\cite{Balkin2021-jr}, it is worth to test this favored regions using the King plot approach itself, rather than relying on other experiments.
As discussed in Refs.~\cite{Frugiuele2017-mp, Ishiyama2026-wd, Fuchs2025-fs}, the favored regions can be tested by a simultaneous fit of multiple transition data sets. 
In particular, Ref.~\cite{Ishiyama2026-wd} suggested that the precision IS measurements for at least five transitions are required to test this assumption for arbitrary masses of the new particle in the Yb King plot, which motivates us to measure ISs of another clock transition.

\begin{figure*}[!hbt]
    \centering
    \includegraphics[width=0.98\linewidth]{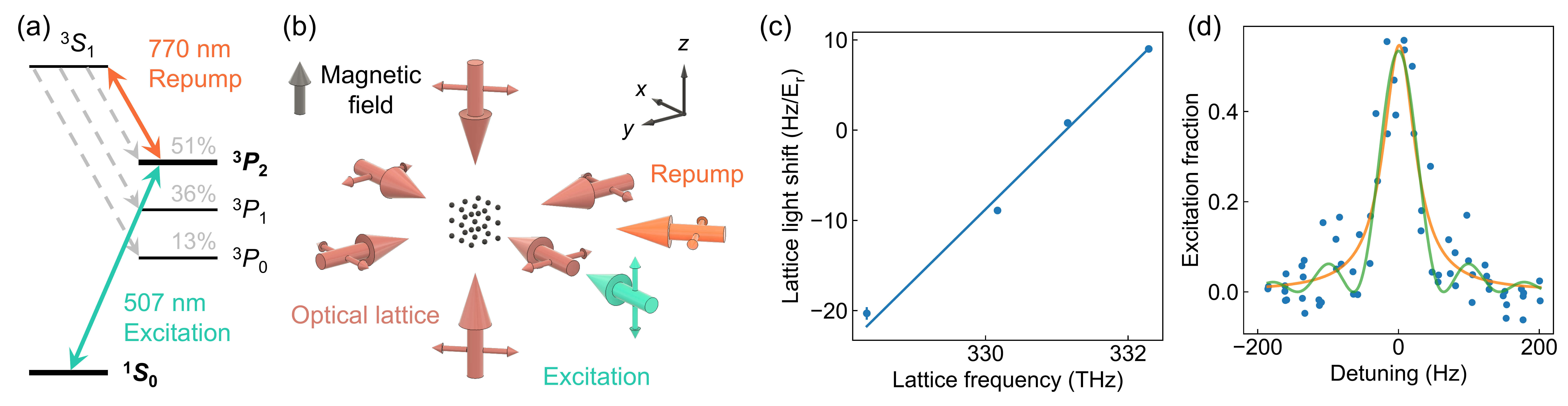}
    \caption{Experimental setup and precision spectroscopy. (a) Energy structure of Yb relevant to our experiments. Note that transitions for laser cooling, imaging, and photo-association are not shown here. Branching ratios for the radiative decay from the $^3S_1$ state are shown in gray~\cite{Porsev1999-ja}. (b) Laser configuration around atoms. The small arrows perpendicular to the main ones represent the laser polarization. (c) Magic wavelength search. The vertical error bars denote the $1\sigma$ statistical uncertainties. The solid line is a linear fit, yielding a magic frequency of $331.13(8)$~THz, corresponding to $905.4(2)$~nm. (d) Excitation spectrum of the $\ket{g} = \g \leftrightarrow \ket{e} = \e \ (m_J = 0)$ transition in $^{174}$Yb with a pulse duration of $13.5$ ms. The green solid curve is a fit with the Rabi spectrum, where a Fourier-limited linewidth is 59 Hz and a maximum excitation fraction is determined as $0.53(2)$. The orange solid curve represents a Lorentzian fit, yielding a full width at half maximum of $55(5)$~Hz and a maximum excitation fraction of $0.55(4)$.}
    \label{fig: spectroscopy}
\end{figure*}

The $\clock$ transition in neutral Yb is an ideal candidate for these purposes.
The excited $\e$ state has a radiative lifetime of about $15$~s, and the transition from the ground $\g$ state at $507$~nm is weakly allowed by a magnetic-quadrupole transition mechanism~\cite{Migdalek1991-hr,Porsev1999-ja,Mishra2001-rj,Dzuba2018-kc,Tang2023-ju}.
Since the first direct excitation~\cite{Yamaguchi2010-va}, this metastable state and the clock transition have been extensively investigated~\cite{Yamaguchi2008-pz, Uetake2012-ph, Kato2013-jz, Taie2016-ur, Konishi2016-xn, Takasu2017-it, Schafer2017-qx, Schafer2017-sp}.
Moreover, they have been applied to quantum simulation~\cite{Kato2016-ut, Tomita2019-td, Takasu2020-nr, Honda2025-yp} and quantum computation~\cite{Okuno2022-cl, Nakamura2024-gu, Kusano2025-ms}, while similar studies using the $\clock$ transition in strontium have also been carried out~\cite{Pucher2024-of, Unnikrishnan2024-hc, Ammenwerth2025-km, Tao2025-od}.
In spite of these works, precision spectroscopy has not been demonstrated for Yb; the narrowest linewidth of atomic spectra so far was about 1 kHz, limited by a finite laser linewidth and the inhomogeneity of the light shift induced by non-magic-wavelength trapping light~\cite{Konishi2016-xn, Kato2016-ut}.
To achieve an accuracy comparable to those of other clock transitions~\cite{Ono2022-oy, Ishiyama2026-wd, Door2025-manual}, the demonstration of precision spectroscopy in a magic wavelength optical lattice is crucial.
Moreover, this would also be useful for quantum technologies, such as high-fidelity two-qubit gates~\cite{Madjarov2020-pu}.

In this Letter, we present precision spectroscopy and IS measurements of the $\ket{g} = \g \leftrightarrow \ket{e} = \e \ (m_J = 0)$ transition.
The experimental identification of a magic wavelength at 905.4(2) nm enables precision spectroscopy with a linewidth well below 100 Hz.
We apply an interleaved clock operation between two isotopes~\cite{Ono2022-oy, Ishiyama2026-wd} and successfully measure ISs of four isotope pairs with Hz-level precision.
Leveraging the acquired precision data together with those of other precisely measured transitions~\cite{Ono2022-oy, Ishiyama2026-wd, Door2025-manual}, we perform a simultaneous fit of the 3D King plot with new bosons~\cite{Ishiyama2026-wd}.
Our analysis reveals that the restricted ansatz, comprising the leading mass and field shifts, one effective factorized higher-order SM contribution, and the new boson, is inconsistent with the possibility of attributing the observed nonlinearity of the 3D King plot solely to the new boson over the mass regions of $m_\phi \leq 6.5$~keV and 18~keV $\leq m_\phi <$ 100 keV, even when considering theoretical uncertainties of the new-physics sensitivity factors.

\paragraph*{Precision spectroscopy.}

Energy structure relevant to our experiments and the schematic of the setup are shown in Figs.~\ref{fig: spectroscopy} (a) and (b), respectively.
We basically employ a similar setup as in our previous works~\cite{Ono2022-oy, Ishiyama2026-wd} except for the excitation and lattice wavelengths.
Ultracold Yb atoms are prepared in a 3D optical lattice, followed by the illumination of a photo-association light to remove multiply-occupied sites.
The excitation laser at 507 nm is then irradiated along the $x$-axis lattice, where a magnetic field of $14.6$~mT is applied along the $z$-axis to utilize the magnetic field-induced electric-dipole transition mechanism~\cite{Taichenachev2006-cu}.
The 507-nm laser is frequency stabilized to an ultra-stable 578-nm cavity made of ultralow expansion glass using the stability transfer method via a frequency comb~\cite{Yamaguchi2012-zf} (see Supplemental Material~\cite{suppl_ref} for the details).
\nocite{Takata2019-hj, Black2001-bi, Ma1994-bh, Ushijima2018-qc, Angeli2013-pc, Kawasaki2024-co, Wang2021-ar, Delaunay2017PRD, Pople1968JCP_FFPT, Norman2018_property, Kuroda2022PRA, Fornberg1988MC_FFPT, Knecht2011TCA_Mossbauer, Knecht_thesis, saue2020dirac, DIRAC22, Sikkema_Visscher_Saue_Ilias_2009, visscher1997ADNDT, dyall1989grasp, Dyall_4f, shee2018equation, Fleig2003JCP, Knecht2010JCP, Thyssen_thesis, Bellafont2015JCP_Delta_Koopmans, South2016PCCP, Hirao2025JPCA_core}
The excitation fraction is determined by the absorption imaging with the $\g \leftrightarrow {}^1P_1$ transition, combined with repumping of the excited state with the $\e \leftrightarrow {}^3S_1$ transition, where the experimentally observed population in the ground state is divided by the imperfect repumping efficiency of 87\% estimated from the theoretical calculation~\cite{Porsev1999-ja}.

A magic wavelength is crucially important to perform precision spectroscopy, at which the polarizabilities of the ground and excited state coincide and the transition frequency between them is free from the perturbation by the trapping light~\cite{Katori2003-hb}.
A magic wavelength is theoretically estimated to exist at 908 nm~\cite{Dzuba2018-kc} or 901(10) nm~\cite{Tang2023-ju} for the condition of $m_J = 0$ and $\theta=\pi/2$, where $\theta$ is the relative angle between the quantization axis and the laser polarization.
Figure~\ref{fig: spectroscopy}(c) shows the frequency dependence of the lattice light shift induced by the $y$-axis lattice after normalization by the $y$-axis lattice depth.
Here, $E_r = h \times 1.4$~kHz is the recoil energy of $^{174}$Yb at 905.4 nm, and $h$ is the Planck constant.
From a linear fit, the magic wavelength is determined as $905.4(2)$~nm, which is reasonably consistent with the theoretical calculations~\cite{Dzuba2018-kc,Tang2023-ju}.

Figure~\ref{fig: spectroscopy}(d) represents typical $\ket{g}  \leftrightarrow \ket{e}$ excitation spectrum of ${}^{174}$Yb with a pulse duration of $13.5$ ms.
The lattice frequency is set to be the determined magic condition, while the lattice depths are $30 \ E_r$ for all three axes.
We fit the data with the function assuming the coherent Rabi excitation (the green solid line), as well as the Lorentzian function (the orange solid line).
The observed linewidth is about 60~Hz, two orders of magnitude narrower than those in the previous studies~\cite{Konishi2016-xn, Kato2016-ut}.

\paragraph*{Isotope shift measurements.}

\begin{figure}[!t]
    \centering
    \includegraphics[width=0.85\linewidth]{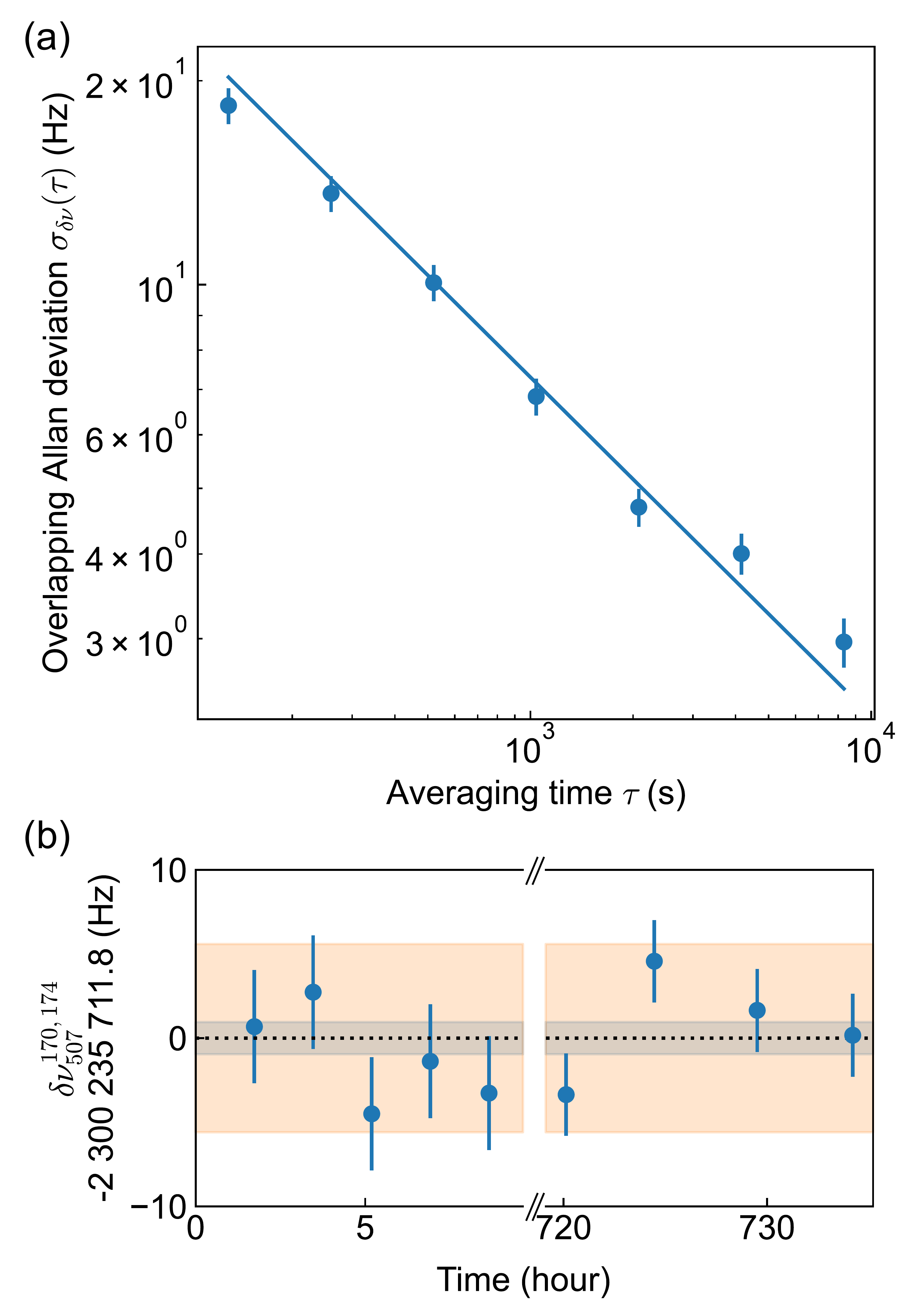}
    \caption{IS measurements. (a) Stability of $\dn{507}{170, 174}$ measurements. The solid line is a fitting curve assuming the white frequency noise, yielding $234(9) \ \mathrm{Hz} / \sqrt{\tau \ \mathrm{(s)}}$. (b) The time trace of $\dn{507}{170, 174}$ measurements. The left and right datasets correspond to two distinct measurements performed on different days. Note that there is no data point during the time interval between the two time durations. The error bar is the $1\sigma$ statistical uncertainty determined from the overlapping Allan deviation. The blue (orange) shaded area represents the $1\sigma$ statistical (systematic) uncertainty of the entire data set.}
    \label{fig: IS}
\end{figure}

Leveraging the high precision of our spectroscopy, we determine the ISs among five stable bosonic isotopes, $\dn{507}{A'A} = \nu_{507}^{A'} - \nu_{507}^A$, where $\nu_{507}^A$ is the transition frequency of the isotope ${}^A$Yb.
We carry out an interleaved clock operation, in which isotope-common perturbations are minimized~\cite{Takano2017-hz, Ono2022-oy, Ishiyama2026-wd}.
Figure~\ref{fig: IS}(a) shows the stability of the $\dn{507}{170, 174}$ measurements.
The overlapping Allan deviation can be fitted well with a curve assuming a white frequency noise, indicating that our measurements do not suffer from any serious long-term perturbation.
Furthermore, to ensure reproducibility, we confirm that two distinct measurements on different days are consistent, as presented in Fig.~\ref{fig: IS}(b).
Table~\ref{tab: IS summary} displays the summary of IS measurements for all four isotope pairs.
As a result, we successfully determine the ISs with Hz-level uncertainties, with a precision similar to those in previous studies~\cite{Ono2022-oy,Door2025-manual,Ishiyama2026-wd} (see Supplemental Material~\cite{suppl_ref} for the details).

\begin{table}[!t]
    \caption{Summary of measured ISs. Total $1\sigma$ uncertainties are shown in parentheses.}
    \begin{ruledtabular}
    \label{tab: IS summary}
    \begin{tabular}{cd}
    Isotope pair $(A', A)$ & \multicolumn{1}{c}{IS $\dn{507}{A'A}$ (Hz)}\\
    \colrule
    $(168, 174)$ & 3~677~546~251.3(6.8) \\
    $(170, 174)$ & 2~300~235~711.8(5.7) \\
    $(172, 174)$ & 1~006~735~627.8(4.9) \\
    $(176, 174)$ &  -960~303~320.6(7.0)  \\
    \end{tabular}
    \end{ruledtabular}
\end{table}

\paragraph*{King plot analysis.}
\begin{figure*}[!ht]
    \centering
    \includegraphics[width=0.98\linewidth]{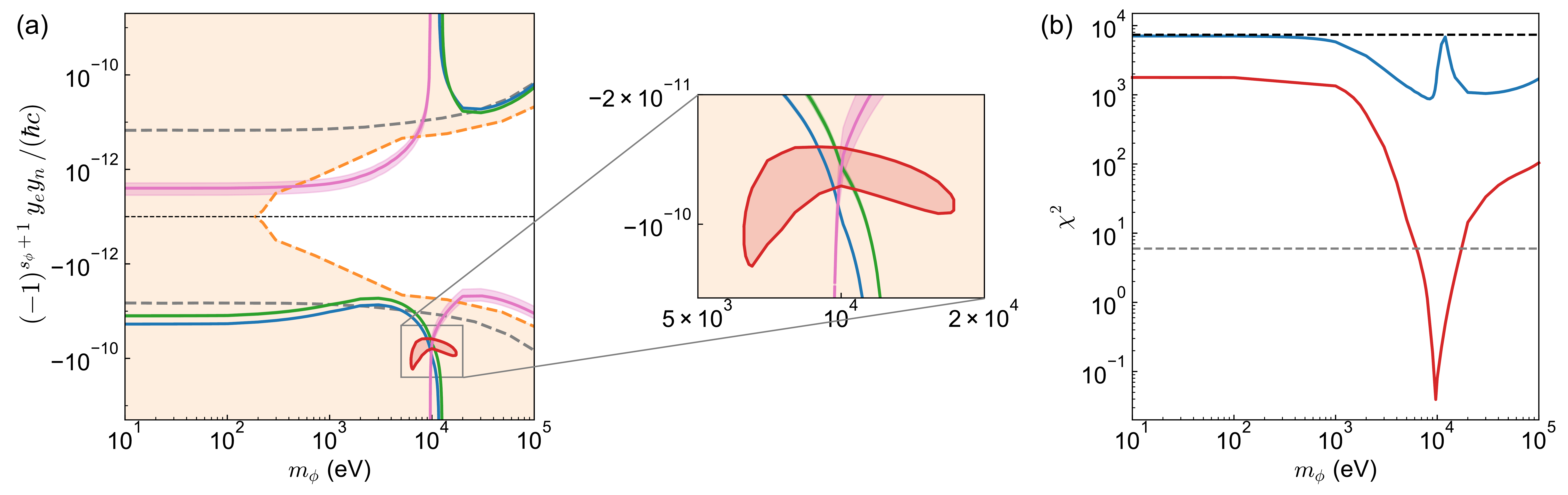}
    \caption{King plot analysis. (a) Product of couplings $y_e y_n$ of a new boson as a function of the mass $m_\phi$. The blue, green, and magenta lines represent the central values of $y_e y_n$ for (411, 431, 578), (411, 467, 578), and (467, 507, 578), respectively, and the corresponding shaded areas display the fitting errors at the 95\% confidence level. Note that the shaded regions of the former two transition sets are narrower than the width of the solid lines. The red solid line and shaded area depict the 95\% confidence region obtained from the simultaneous fit of the five transitions including $X_i$ uncertainties, which is magnified in the inset. The orange dashed curve is the current best laboratory-based bound, determined as the product of the constraint on $y_e$ from a $(g-2)_e$ measurement~\cite{Fan2023-cz} and the constraint on $y_n$ from neutron scattering measurements~\cite{Leeb1992-vi, Pokotilovski2006-ct, Nesvizhevsky2008-mf}, while the orange shaded area is the corresponding excluded region. The upper bound from the King plot analysis of calcium ions~\cite{Wilzewski2025-dx} is displayed as a gray dashed line. We note that this figure focuses exclusively on laboratory-based experiments, and constraints derived from astrophysical observations~\cite{Berengut2020-mx} are not shown.
    (b) $m_\phi$ dependence of $\chi^2$ in the simultaneous fit. The blue (red) solid line corresponds to the analysis without (with) $X_i$ uncertainties. The black dashed line shows $\chi^2$ of the 3D generalized King plot before including the PS. As a reference, $\chi^2$ values at the 95\% confidence level with the dof of 2 is shown as the gray dashed line.}
    \label{fig: KP}
\end{figure*}

We employ the acquired high-precision IS data to perform a King plot analysis. 
In general, the IS of the transition $i$, $\Dn{i}$, can be separated into distinct contributions~\cite{Berengut2025-yy,Munro-Laylim2022-ho}:
\begin{align} \label{eq: IS}
    \Dn{i} = &K_i \bm{w} + F_i \Dr + G^{(4)}_i \Drr \\ 
             &+ G^{(2)}_i \Ddr + \cdots + \alpha_{\mathrm{NP}} X_i \DA.  \nonumber
\end{align}
Here, the vector notation means that it is composed of the four isotope pairs.
The first two terms, which represent the leading-order mass shift and the leading-order field shift, respectively, are dominant contributions originating from the SM.
Higher-order SM effects are denoted by $G^{(4)}_i \Drr + G^{(2)}_i \Ddr + \cdots$, representing the next-leading-order Seltzer moment, the quadratic field shift, among others (see Supplemental Material~\cite{suppl_ref} for the details of the SM terms of the IS).
Note that the complete IS may contain several higher-order SM contributions and that nuclear polarization, in particular, need not factorize into independent electronic and nuclear factors.
The final term, called the particle shift (PS), arises from a hypothesized new Yukawa potential acting between electrons and neutrons: $V(r) = \left( -1 \right)^{1+s_\phi} y_e y_n \exp{(-m_{\phi}cr/\hbar)}/(4\pi r)$.
Here, $m_{\phi}$ and $s_{\phi}$ represent the mass and spin of the mediating particle.
The coupling constants between the new particle and the electron (neutron) are denoted $y_{e}$ ($y_{n}$), while the overall coupling parameter $\alpha_{\mathrm{NP}}$ is defined as $(-1)^{1+s_{\phi}}y_{e}y_{n}/(4\pi\hbar c)$.
Each of these terms is known to be factorizable as the product of a transition-dependent electronic factor ($K_i$, $F_i$, $G^{(4)}_i$, $G^{(2)}_i$, and $X_i$) and an isotope-dependent nuclear factor. 
The nuclear factor of the mass shift, $\w{A'A} = m_{172}/m_{A'} - m_{172}/m_A$, is derived from nuclear mass ratios measured at $10^{-12}$ level~\cite{Door2025-manual}, where $m_{A}$ is the nuclear mass of ${}^A$Yb.
The nuclear factor denoted by $\delta \langle r^n \rangle^{A'A}= \langle r^n \rangle^{A'} - \langle r^n \rangle^A$ represents the differential $n$-th nuclear charge moment, and the parameter $\ddr{A'A}$ is defined as $(\dr{A'A_0})^2 - (\dr{AA_0})^2$ relative to a reference isotope $A_0$.
$\DA$ is the difference of the mass number.

First, we present the linearity test of the 3D generalized King plot using a restricted phenomenological ansatz.
Let us assume that the IS consists of three terms, namely, the leading-order mass shift, the leading-order field shift, and one higher-order SM term factorizable into nuclear and electronic factors.
Then, the following relation between three transitions should be satisfied:
\begin{equation}  \label{eq: individual 3D}
    \Dn{3} = k \bm{w} + f_1 \Dn{1} + f_2\Dn{2},
\end{equation}
which is known as the 3D generalized King linearity.
Although Eq.~(\ref{eq: individual 3D}) itself involves only three transitions, it is possible to fit simultaneously four or more transitions~\cite{Frugiuele2017-mp, Ishiyama2026-wd, Fuchs2025-fs}.
The simultaneous fit of three independent 3D King linearities among precisely measured five transitions ($i = 411, \ 431, \ 467, \ 507, \ 578$) results in $\chi^2 = 7383$ with the degrees of freedom (dof) of 3, corresponding to $86\sigma$ (see Supplemental Material~\cite{suppl_ref} for the details).
It is noted that the $86\sigma$ significance quantifies the rejection of the restricted ansatz and should not be interpreted as a significance relative to the complete SM.

The observed huge nonlinearity indicates that the reduced three-term description in Eq.~(\ref{eq: individual 3D}) is insufficient.
The important question is whether the additional term can be attributed to the PS or not.
Let us suppose that the IS is composed of the three terms assumed above and the PS.
Then, the restricted phenomenological ansatz in Eq.~(\ref{eq: individual 3D}) is modified as
\begin{eqnarray}  \label{eq: individual 3D PS}
    \Dn{3} &=& k \bm{w} + f_1 \Dn{1} + f_2\Dn{2}\\
    &+& \alpha_{\mathrm{NP}} (X_3 - f_1 X_1 - f_2 X_2) \DA , \nonumber
\end{eqnarray}
where $\alpha_{\mathrm{NP}}$ is obtained from a fit by this formula with theory inputs of $X_i$ (see Supplemental Material~\cite{suppl_ref} for the computational details).
If the observed nonlinearity is solely attributed to the new particle, the coupling constants $y_e y_n$ should be independent of the set of the transitions.
However, as shown in Fig.~\ref{fig: KP}(a), the favored regions of the new physics parameters are different among the combinations, meaning that the assumption is not reasonable.

To quantify the inconsistency among the favored regions shown in Fig.~\ref{fig: KP}(a), we consider another analysis based on the combined King relation~\cite{Ishiyama2026-wd}.
The blue solid line in Fig.~\ref{fig: KP}(b) shows the $m_\phi$ dependence of $\chi^2$, where all five transitions are fitted by Eq.~(\ref{eq: individual 3D PS}) with one fitting parameter for $\alpha_\mathrm{NP}$.
Significant $\chi^2$ persists for arbitrary $m_\phi$; as much as 872 of $\chi^2$ still remains, even at the minimum point $m_\phi = 8.40$~keV, which excludes the possibility of attributing the observed nonlinearity of the 3D King plot solely to the new physics within the restricted ansatz (see Supplemental Material~\cite{suppl_ref} for the details).
Here, to be more conservative, we take the uncertainty of $X_i$ into account~\cite{Hur2022-manual,Wilzewski2025-dx, Fuchs2025-fs}, shown as the red solid line in Fig.~\ref{fig: KP}(b) (see the Supplemental Material~\cite{suppl_ref} for the evaluation of $X_i$ uncertainties).
We find a sizable $\chi^2$ for a given mass in the small-mass region of $m_\phi \leq 6.5$~keV, as well as 18~keV $\leq m_\phi <$ 100~keV, indicating that the PS is not the main source of the nonlinearity observed in the 3D generalized King plot.
Note that the region of $m_\phi \geq 100$~keV is not considered in this Letter, since the PS effects are indistinguishable from the leading- and/or higher-order field shifts~\cite{Berengut2018-su}.
It is also noted that the included $X_i$ uncertainties concern the calculated new-particle sensitivity factors and do not represent uncertainties associated with the omitted higher-order SM contributions.

The $\chi^2$ value reaches its minimum of 0.04 at $m_\phi = 9.67$~keV.
In this area, the existence of a new particle cannot be ruled out based on the current Yb data sets.
This is, however, entirely due to the lack of the new particle sensitivities around this mass region, originating from the specific property of the electronic wavefunctions of Yb atoms.
The increased dof, resulting from the new $\clock$ IS data, allows us to include the new particle mass $m_{\phi}$ as an additional fitting parameter, alongside its coupling constants $y_e y_n$~\cite{Ishiyama2026-wd}.
As a result, we obtain the 95\% confidence region in the $(m_\phi, \ y_e y_n)$ space, as shown by the red shaded area in Fig.~\ref{fig: KP}(a), which roughly coincides with the intersection of the three favored regions (see Supplemental Material~\cite{suppl_ref} for the details).
It is noted again that the low $\chi^2$ near $m_\phi=9.67$~eV, represented by the red shaded area is due to the loss of discriminatory power within the restricted model in Eg.~(\ref{eq: individual 3D PS}), rather than as a physical favored or allowed region.
Conversely, the poor fits elsewhere demonstrate that the PS term alone does not restore consistency within that model; they do not determine the relative size of a possible particle contribution.
We also note that the upper bound by the IS data of Ca ions shown by the gray dashed lines in Fig.~\ref{fig: KP}(a) seems disfavor the allowed region in the present Yb IS analysis.

Finally, we address the favored region obtained for the transition set (467, 507, 578), as shown by the magenta line and shaded region, which lies outside the existing excluded parameter space.
In general, an analysis based on Eq.~(\ref{eq: individual 3D PS}) yields a small coupling constant when a King plot exhibits a small nonlinearity.
Naively, this region might be interpreted as suggesting the existence of the new particle.
However, the underlying assumption in Eq.~(\ref{eq: individual 3D PS}) --- that the IS is composed of the three SM terms and the PS --- is excluded by our simultaneous fit.
We emphasize that focusing only on a specific transition set that does not conflict with the existing bound may lead to an incorrect conclusion, which demonstrates the importance of our simultaneous fit analysis using five transitions with the new one of the $\clock$ transition.

\paragraph*{Summary.}
In this Letter, we report ultra-narrow spectroscopy of the $\clock$ transition in Yb, enabled by trapping atoms in a 3D optical lattice at the magic wavelength of $905.4(2)$ nm.
We successfully measure ISs between five bosonic isotopes with Hz-level accuracy, comparable to the other transitions precisely measured so far~\cite{Ono2022-oy, Door2025-manual, Ishiyama2026-wd}.
These precise IS data are fully exploited for the King plot analysis.
Our analysis reveals that, within the restricted ansatz comprising the leading mass and field shifts, one effective factorized higher-order SM contribution, and the PS, the five-transition data are inconsistent with the PS serving as the sole additional contribution over the mass ranges of $m_\phi \leq 6.5$~keV and 18~keV $\leq m_\phi <$ 100 keV. This result does not exclude a boson contribution in the presence of further higher-order SM effects.

To exclude the possibility of new-physics interpretation around 9.67~keV, clock transitions with very different electronic structures, such as those in highly-charged ions, might be helpful~\cite{Schauer2010-th, Staiger2025-lo-manual}.
To get information about the new particle in the presence of the higher-order SM effect in the 3D King plot, the SM subtraction method using electronic factor ratios and/or the construction of a 4D King plot with a radio-active isotope $^{166}$Yb are promising.
In both cases, it is vital to correctly account for higher-order SM terms.
Of particular note is the nuclear polarization; its possibly non-factorizable nature might prevent us from eliminating it simply by expanding the dimensionality of the generalized King plot.
In this regard, the development of advanced atomic and nuclear many-body theory will be indispensable for future King plot analyses~\cite{Flambaum2018-bm,Flambaum2021-th,Munro-Laylim2022-ho,Viatkina2023-lm,Assi2025-jz-manual}.
Finally, the analysis method employed in this work enables the effective use of multiple transitions, thereby encouraging us to perform precision IS measurements on a larger number of transitions, extending beyond Yb.


\paragraph*{Acknowledgments.}
We thank Amar Vutha, Yosuke Takasu, Naoya Ozawa, and Yushiro Nishikawa for useful discussions.
This work was supported by
the Grant-in-Aid for Scientific Research of JSPS (No. JP22K20356, No. 24K16995, No. 24KJ1347, No. 24K07018), 
JST PRESTO (No. JP-MJPR23F5), 
the Matsuo Foundation, 
JST CREST (Nos. JPMJCR1673 and JP-MJCR23I3), 
MEXT Quantum Leap Flagship Program (MEXT Q-LEAP) Grant No. JPMXS0118069021, 
JST Moon-shot R\&D (Grant No. JPMJMS2268 and JP-MJMS2269), 
JST ASPIRE (No. JPMJAP24C2), 
and the RIKEN TRIP initiative.
We used the supercomputer of ACCMS, Kyoto University (Service Course and Collaborative Research Project for Enhancing Performance of Programming). 

\paragraph*{Data availability.}
There are no publicly available research data or software supporting this manuscript. Requests for further information or data should be sent to the authors.


%

\onecolumngrid
\section*{End Matter}
\twocolumngrid

\paragraph*{Magic wavelength.}
We experimentally search the magic wavelength around these wavelengths using a widely tunable titanium-sapphire laser, in which we vary the $y$-axis lattice depth with 15, 25 and 35 $E_r$ in an interleaved manner, while those of the other two axes is set constant as 30 $E_r$.
The lattice depth is calibrated using a pulsed lattice technique~\cite{Denschlag2002-hb}, where the uncertainty of the lattice depth is estimated to be 7.9\%.
As a result, we obtain the differential light shift normalized by the $y$-axis lattice depth, in units of Hz/$E_r$, as shown Fig.~\ref{fig: spectroscopy}(c).

The blue solid line in Fig.~\ref{fig: spectroscopy}(c) represents a linear fit to the measured lattice light shift data, showing the nonlinearity corresponding to the reduced $\chi^2$ ($\chi^2_\mathrm{red}$) of 10.6.
This relatively large value is mainly due to the limited number of measurement cycles at each lattice frequency.
To account for this, the $1\sigma$ statistical uncertainty for the magic wavelength is inflated by $\sqrt{\chi^2_\mathrm{red}}$.
Consequently, the magic wavelength is determined as $905.4(2)$~nm.

\paragraph*{Limiting factor of the observed linewidth}
The typical linewidth of the excitation spectra of the $\ket{g} \leftrightarrow \ket{e}$ transition is about 60~Hz, as shown in Fig.~\ref{fig: spectroscopy}(d).
We confirm that the linewidth of the atomic spectrum cannot be further narrowed.
In fact, we find that the linewidth and the non-unity maximum excitation fraction are similar to those observed for another clock transition $\g \leftrightarrow 4f^{13}5d6s^2 \ (J=2, \ m_J=0)$ in our previous work~\cite{Ishiyama2026-wd}.
Given that the excitation lasers in both experiments are stabilized to the same frequency comb referenced to the common 578-nm cavity (see Supplemental Material~\cite{suppl_ref}), this consistency suggests that the common phase noise of the shared reference system is the limiting factor.

\end{document}


\title{Precision Isotope-Shift Measurement of the $\clock$ Transition and \\a Multi-Transition King-Plot Consistency Test - Supplemental Material}
\maketitle


\section{Experimental setup}

Ultracold Yb atoms are prepared in the same manner as in our previous work~\cite{Ishiyama2026-wd}.
A titanium-sapphire laser is employed as an optical lattice, together with a volume Bragg grating to suppress amplified spontaneous emission.
The lattice depth is calibrated using a pulsed lattice technique~\cite{Denschlag2002-hb}, and the laser power is stabilized with acousto-optic modulators (AOMs).

\begin{figure*}[!ht]
    \centering
    \includegraphics[width=0.8\linewidth]{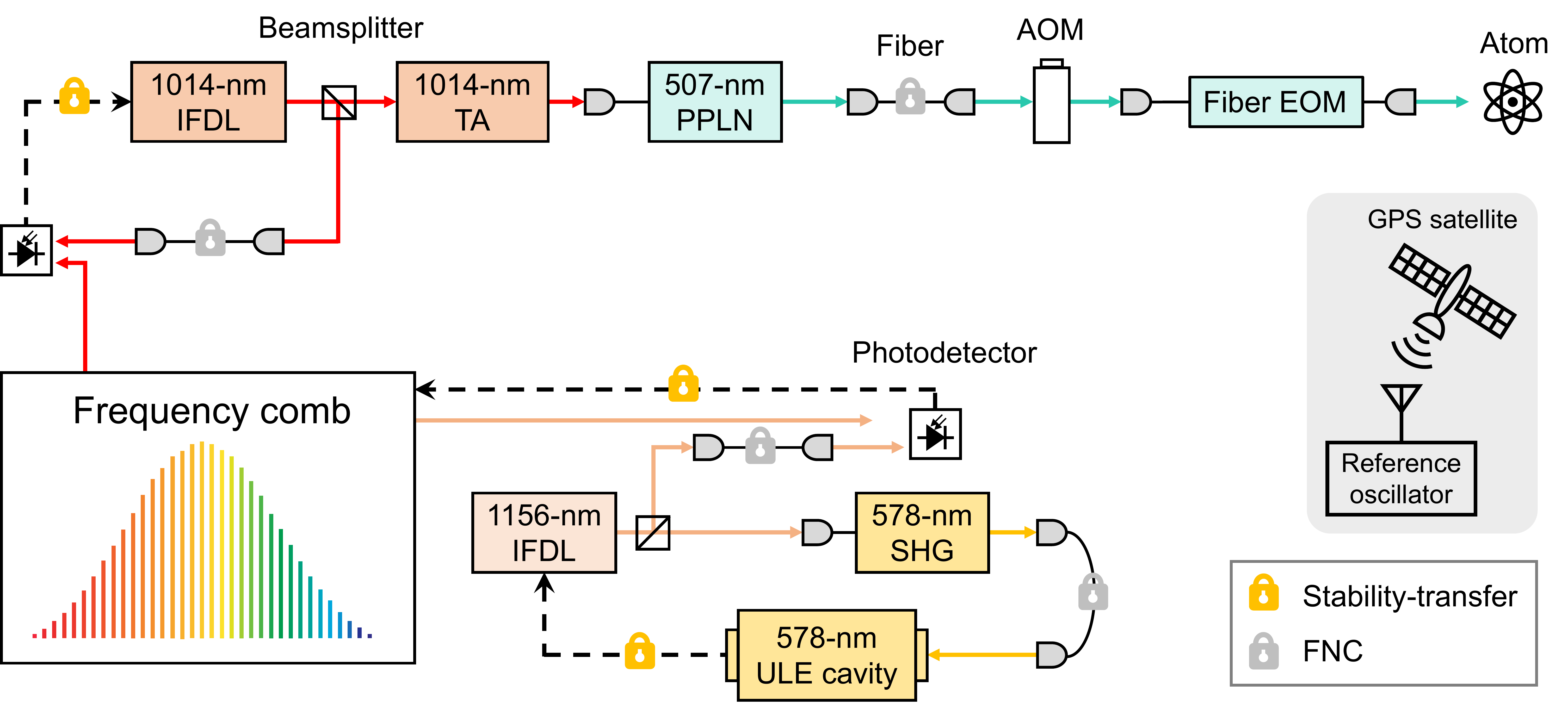}
    \caption{Details of the excitation laser system. The locks of the stability-transfer are illustrated as black dashed arrows.}
    \label{fig: clock laser}
\end{figure*}

Figure~\ref{fig: clock laser} illustrates our excitation laser system.
In this Letter, we extend our previous work~\cite{Ishiyama2026-wd} to 507 nm. 
The repetition rate of the comb is stabilized using a 1156-nm interference-filter-stabilized external-cavity diode laser (IFDL)~\cite{Takata2019-hj}, which is, after second-harmonic-generation (SHG), locked to the 578-nm cavity made of ultralow expansion (ULE) glass with the Pound-Drever-Hall method~\cite{Black2001-bi}.
The carrier-envelope offset frequency is stabilized to a 10-MHz reference oscillator using an $f$-$2f$ interferometer method.
The stability of the comb is then transferred to a 1014-nm IFDL by offset-locking it to the comb.
Note that all the radio-frequency generators relevant to the excitation lasers are referenced to the frequency standard in the National Metrology Institute of Japan via GPS common-view method.
The fiber-noise-cancellation (FNC) system~\cite{Ma1994-bh} is installed for all the optical fibers longer than three meters. 
The 1014-nm laser is amplified with a tapered amplifier (TA), and the wavelength is converted to 507 nm with a commercially available wave-guided periodically poled lithium niobate (PPLN) SHG crystal.
The AOM is used to stabilize a fiber-coupled electro-optic modulator (EOM)'s output power and to switch the laser on and off.
The frequency of the 507-nm laser is fine-tuned with the fiber EOM to perform precision spectroscopy and interleaved clock operation.
The total intensity at the atomic position, including the carrier and sidebands, amounts to $8$ W/cm$^2$.

\section{Isotope shift measurements}
\subsection{Interleaved clock operation}
\begin{figure}[!ht]
    \centering
    \includegraphics[width=0.95\linewidth]{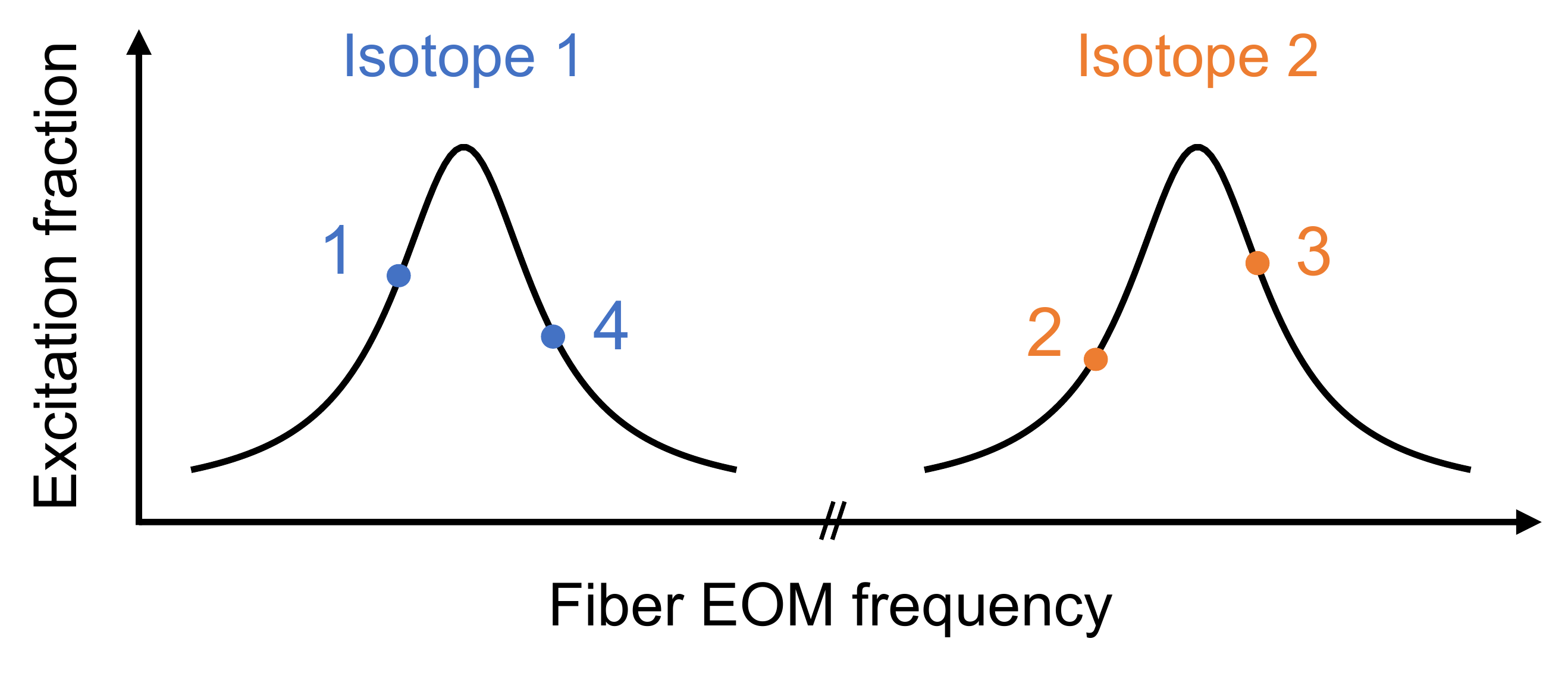}
    \caption{Schematic of the interleaved clock operation. We measure the excitation fractions of four points, and the obtained error signal is fed back to the fiber EOM at 507 nm. The numbers represent the order of the interrogation.}
    \label{fig: clock operation schematic}
\end{figure}

We carry out an interleaved clock operation, where almost all the systematic effects common between isotopes can be mitigated, such as the drifts of the clock-laser frequency, the lattice depth, the excitation laser intensity, and the magnetic field~\cite{Takano2017-hz, Ono2022-oy, Ishiyama2026-wd}.
In this work, the feedback control of the frequency shifter is different from that in our previous work~\cite{Ishiyama2026-wd}.
We measure the excitation fraction at the left and right shoulders for both isotopes, as shown in Fig.~\ref{fig: clock operation schematic}.
The frequency detuning between both shoulders is set to be the full-width at half-maxima of the atomic spectrum, so that the sensitivity of the excitation fraction to the laser phase noise is maximized.
The error signal is defined as the difference between the excitation fractions for the right and left detunings, and fed back to the fiber EOM with the proportional servo method.
To evaluate the linear drift, the resulting time series of the fiber EOM frequencies of both isotopes are simultaneously fitted with linear functions in real time during the measurement campaign, where only the last 20-40 data points are used for the fitting.
The IS is straightforwardly determined as the difference of the center fiber EOM frequencies between two isotopes.

\subsection{Evaluation of statistical uncertainties}
\begin{figure}[!ht]
    \centering
    \includegraphics[width=0.95\linewidth]{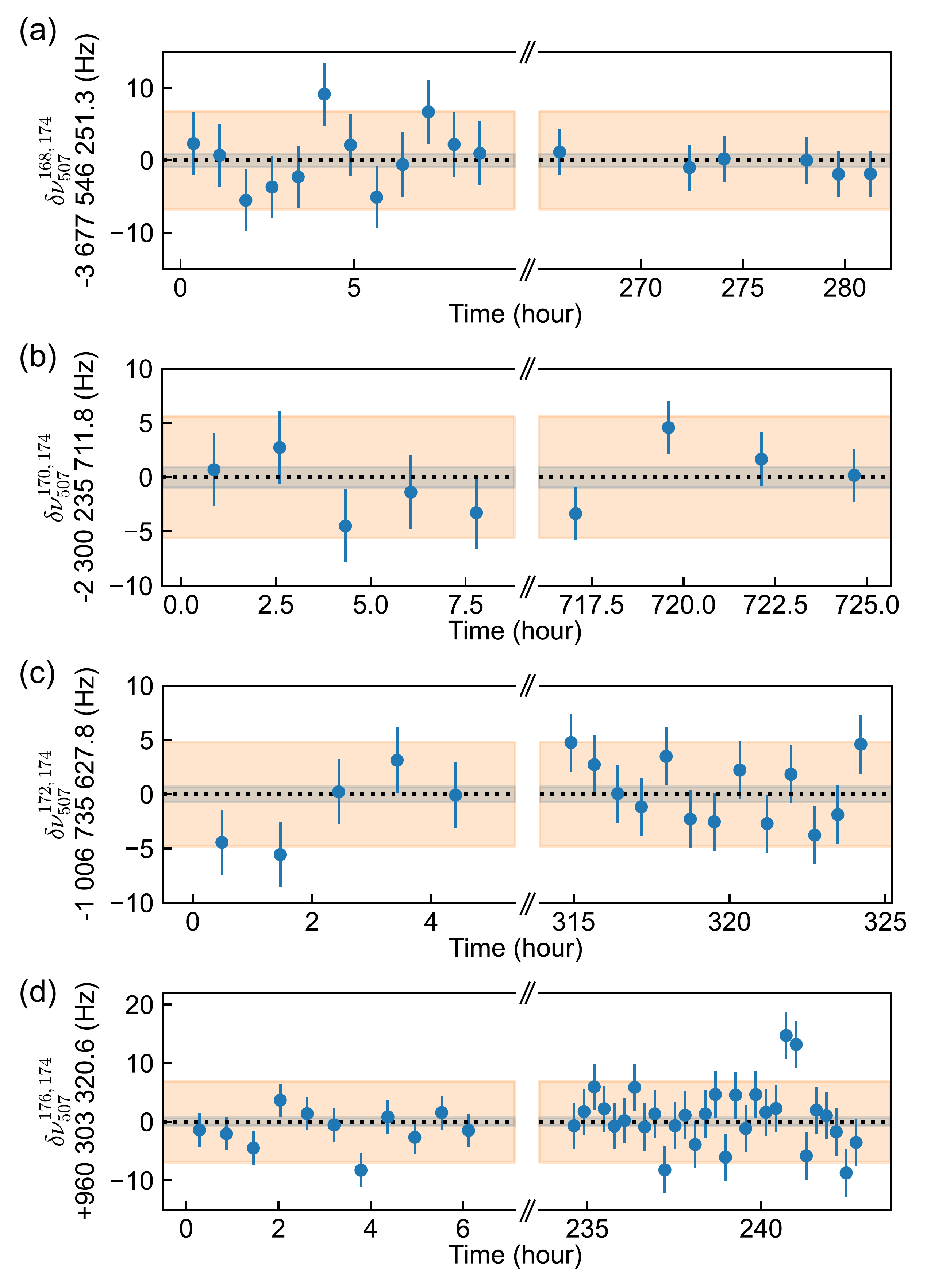}
    \caption{Time traces of IS measurements. Two time durations in each of (a), (b), (c), and (d) correspond to two distinct measurements performed on different days. Note that there is no data point during the time interval between the two time durations. The error bar is the $1\sigma$ statistical uncertainty determined from the overlapping Allan deviation. The blue (orange) shaded area represents the $1\sigma$ statistical (systematic) uncertainty. }
    \label{fig: IS timetrace}
\end{figure}

The clock operation is executed under the standard experimental conditions: the lattice depth (30, 30, 30)~$E_r$, the magnetic field 14.6~mT, and the excitation laser intensity $8$ W/cm$^2$.
For every isotope pair, the data acquisition is repeated on two distinct days to ensure reproducibility. 
The time traces of these IS measurements are illustrated in Fig.~\ref{fig: IS timetrace}.
For each measurement campaign, the overlapping Allan deviation of the interleaved clock operation is analyzed, as depicted in Fig.~2 of the main text, to determine the appropriate averaging time where the deviation aligns with white frequency noise.
The entire dataset is partitioned into segments, and the mean value of each segment, after correction for all systematic effects, is displayed as a blue data point in Fig.~\ref{fig: IS timetrace}. 
The corresponding error bars are set by the overlapping Allan deviation at the previously determined averaging time. 
The final IS value is computed as the weighted average across all segments. 
The $1 \sigma$ statistical uncertainty (represented by the blue shaded area) is evaluated as the maximum value between the weighted standard error and the propagated error of the segment error bars. 
Furthermore, the overall $1\sigma$ systematic uncertainty is presented (orange shaded regions; see Table.~\ref{tab: IS budget}), confirming that the two independent measurements are consistent within the determined $1\sigma$ systematic uncertainties for all isotope pairs.

\subsection{Investigation of systematic effects}
Table~\ref{tab: IS budget} shows the error budget of our IS measurements.
All systematic effects considered in this work will be discussed in the following.

\begin{figure}[!ht]
    \centering
    \includegraphics[width=0.95\linewidth]{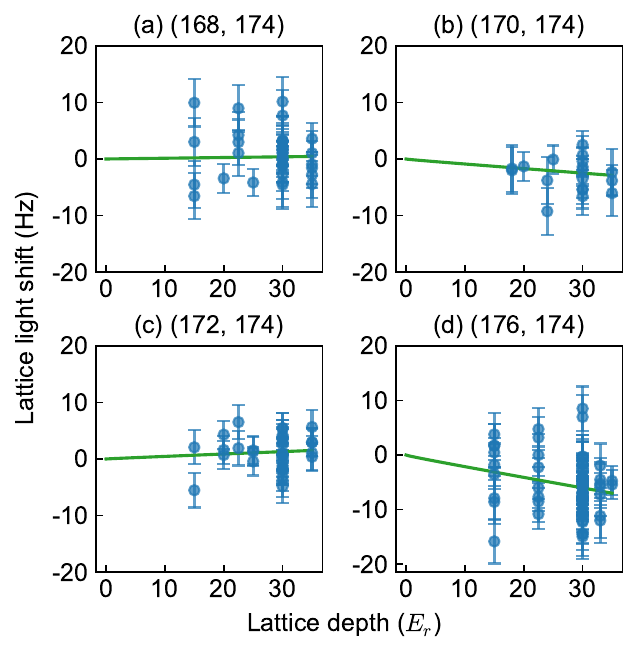}
    \caption{Lattice light shifts. The measured ISs $\dn{507}{A, 174}$ are plotted with respect to the lattice depth per axis. Error bars represent the $1\sigma$ statistical uncertainties obtained from the overlapping Allan deviations. Green solid lines are fitting curves by Eq.~(\ref{eq: LLS}) and the determined $y$-axis intercepts are subtracted from the data.}
    \label{fig: IS LLS}
\end{figure}

\textit{Lattice light shift.}
Since the magic wavelength is isotope-dependent in principle, the lattice light shift (LLS) should be carefully evaluated.
We investigate the lattice depth dependence of the ISs, as summarized in Fig.~\ref{fig: IS LLS}. 
Note that we keep the lattice depth ratio for the three axes, since the excited state $\e$ has a tensor light shift, which might cause the axis-dependent light shift.
To satisfy the Lamb-Dicke condition, the lattice depth is kept more than $15 \ E_r$, corresponding to the Lamb-Dicke factor of $0.64$.
    We determine the LLS at the operational lattice depth of $30 \ E_r$ by fitting the experimental data using the following expression, which accounts for the nonlinear effect arising from the atomic zero-point energy~\cite{Ushijima2018-qc}:
\begin{equation} \label{eq: LLS}
    f(s) = a \left( s + \frac{\sqrt{s}}{2} \right) + b.
\end{equation}
Here, $s$ is the lattice depth expressed in units of recoil energy $E_r$, and $a$ and $b$ are the resultant fitting coefficients. 
Note that we assume the isotope dependence of any potential multipolar effects is negligible. 
Uncertainty associated with the lattice light shift correction is evaluated by combining the statistical fitting errors with the estimated $7.9\%$ uncertainty in the lattice depth calibration, where the former gives the dominant contribution.

\begin{table*}[!th]
    \caption{Error budgets of IS measurements in units of Hz. All systematic corrections and $1\sigma$ uncertainties of $\dn{507}{A'A}$ are presented except AOM chirp and BBR shift, which are negligibly small.}
    \begin{ruledtabular}
    \label{tab: IS budget}
    \begin{tabular}{lrrrrrrrr}
    Isotope pair $(A', A)$ & \multicolumn{2}{c}{$(168, 174)$} & \multicolumn{2}{c}{$(170, 174)$} & \multicolumn{2}{c}{$(172, 174)$} & \multicolumn{2}{c}{$(176, 174)$} \\
    \colrule
    Systematic effects & Corr. & Unc. & Corr. & Unc. & Corr. & Unc. & Corr. & Unc. \\
    \cline{1-1} \cline{2-3} \cline{4-5} \cline{6-7} \cline{8-9}
    Lattice light shift & 0.4 & 3.3 & 2.3 & 4.4 & -1.2 & 2.8 & 5.5 & 2.8 \\
    Probe light shift & 0 & 4.1 & 0 & 2.4 & 0 & 3.1 & 0 & 5.7 \\
    Quadratic Zeeman shift & 0 & 4.2 & 0 & 2.5 & 0 & 2.4 & 0 & 2.7 \\
    Servo error & 0.0 & 0.3 & 0.0 & 0.2 & -0.1 & 0.2 & 0.0 & 0.2 \\
    Total & 0.4 & 6.8 & 2.2 & 5.6 & -1.3 & 4.8 & 5.5 & 6.9 \\
    \end{tabular}
    \end{ruledtabular}
\end{table*}

The correction value derived from the aforementioned fitting analysis shows a relatively large value for the $(176, 174)$ pair, which has one of the smallest mass number differences.
However, it is generally expected that the LLS is approximately proportional to the difference in mass numbers between the isotope pairs.
The scattering of the data points can result in such a finite correction value.
To pursue a more robust correction, we consider an alternative method in which the data from all four isotope pairs are simultaneously fitted.
This approach is based on the theoretical expectation that the differential light shift by the optical lattice for each isotope is proportional to the differential polarizability $\Delta \alpha = \alpha_e - \alpha_g$, which is given by
\begin{equation} \label{eq: pol}
    \alpha_i \propto \sum_{j} \frac{k_{i, j}}{\nu - \nu_{i,j}}.
\end{equation}
Here, $j$ spans all states that are coupled to the state $i$ with optical transitions.
$\nu$ is the laser frequency, while $\nu_{i,j}$ represents the transition frequency between the two states.
$k_{i, j}$ is a constant independent of $\nu$ and $\nu_{i,j}$, including the transition matrix element between the two states. 

The isotope dependence of $\Delta \alpha$ primarily arises from the IS of $\nu_{i,j}$.
To set a constraint on the isotope dependence of $a$ in Eq.~(\ref{eq: LLS}), we treat the IS as being given by the leading-order field shift in a crude approximation. 
This assumption is plausible, since the IS is dominated by the leading order field shift for a heavy element such as Yb.
Using the literature values for $\dr{}$~\cite{Angeli2013-pc}, $a$ in Eq.~(\ref{eq: LLS}) for four isotope pairs are theoretically anticipated to satisfy the following relation: $(a^{168, 174}, \ a^{170, 174}, \ a^{172, 174}) = (-3.66, \ -2.32, \ -1.04) \ a^{176, 174}$.
Based on this expected scaling, a simultaneous fit of the four isotope pairs yields a new correction value for the $(176, 174)$ pair, which turns out to be $0.7(9)$ Hz.

The discrepancy between the correction values obtained from the two methods is $4.8(2.9)$~Hz, which is within the total systematic uncertainty 6.9~Hz. 
Consequently, the final IS value is not significantly affected by the choice of the analysis method.
In this Letter, we adhere to the individual fitting approach, same as our previous work~\cite{Ishiyama2026-wd}, since the latter method relies on an approximation that simplifies the IS to only the leading-order field shift, and it is difficult to properly assess the uncertainty of the underlying model. 

\begin{figure}[!t]
    \centering
    \includegraphics[width=0.95\linewidth]{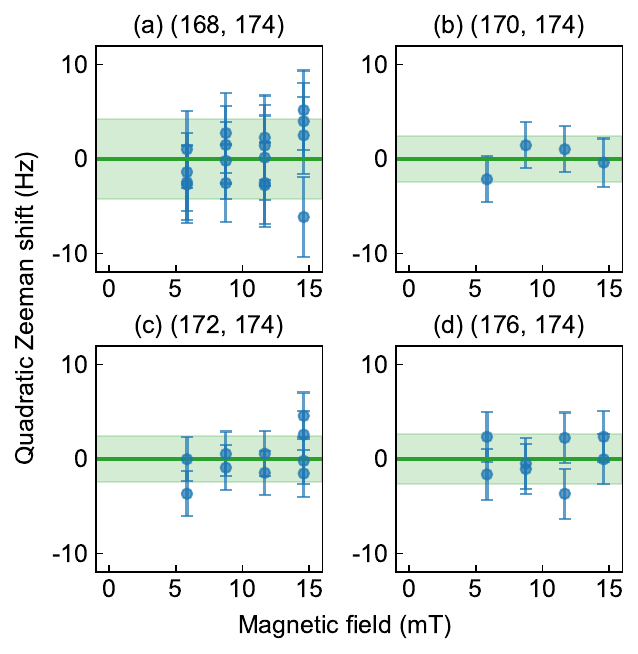}
    \caption{Quadratic Zeeman shifts. The magnetic field dependence of ISs is investigated. The definition of the error bars are the same as in Fig.~\ref{fig: IS LLS}. Green shaded areas depict the $1\sigma$ systematic uncertainties, whose definitions are provided in the text.}
    \label{fig: IS QZS}
\end{figure}

\textit{Quadratic Zeeman shift.}
The first-order Zeeman shift is absent since the upper and lower states in the clock transition are $m_J=0$.
However, the quadratic Zeeman shift (QZS) should be considered.
Theoretical estimates suggest a QZS coefficient of $1.2 \times 10^6 \ \mathrm{Hz/T}^2$ for the $\gtoe$ transition, equating to a shift of approximately $260~\mathrm{Hz}$ at the operational field of $14.6~\mathrm{mT}$~\cite{Dzuba2018-kc}. 
The isotope dependence of the QZS can be theoretically estimated as $10^{-4}$ order of magnitude, well below our measurement precision. 
Experimental confirmation is provided by Fig.~\ref{fig: IS QZS}, showing no significant dependence of the measured ISs on the magnetic field magnitude.
Thus, the correction for the QZS is set to zero.
The $1\sigma$ uncertainty is conservatively determined as the larger of the weighted standard deviation and the average of error bars.

\begin{figure}[!t]
    \centering
    \includegraphics[width=0.95\linewidth]{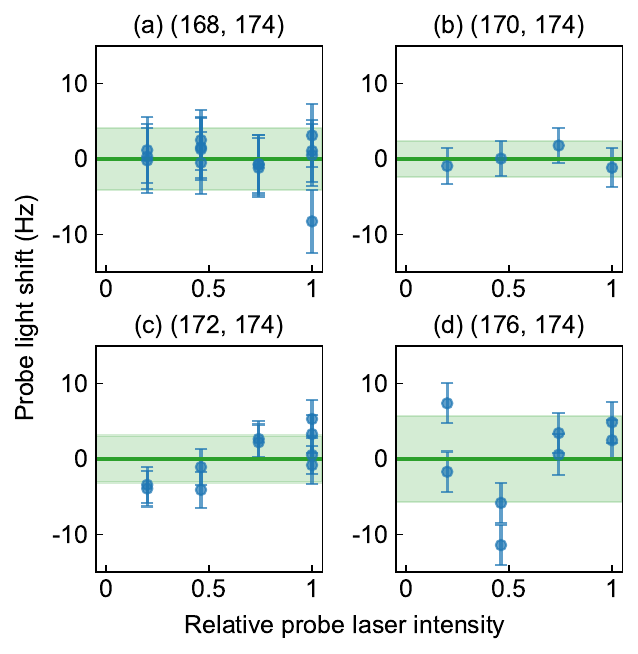}
    \caption{Probe light shifts. The horizontal axis represents the probe laser intensity normalized with that of the standard condition. The definitions of the error bars and green shaded areas are the same as in Fig.~\ref{fig: IS QZS}.}
    \label{fig: IS PLS}
\end{figure}

\textit{Probe light shift.}
The probe light shift (PLS), which arises from the AC Stark shift induced by the clock excitation laser itself, is also anticipated to be nearly isotope-independent, similar to the QZS.
As depicted in Fig.~\ref{fig: IS PLS}, the IS measurements exhibit no systematic dependence on the excitation laser intensity. 
Given this negligible dependence, the correction and uncertainty associated with the PLS are determined in the same manner as for the QZS.

\textit{Servo error.}
The servo error inherent in the clock operation is quantified by analyzing the mean values and the corresponding overlapping Allan deviation of the excitation fraction differences at the two shoulders of the atomic resonance profile.

\textit{Black body radiation shift.}
The Black Body Radiation (BBR) shift is theoretically calculated to be $2.3~\mathrm{Hz}$ at $300~\mathrm{K}$~\cite{Dzuba2018-kc} and is basically anticipated to act as a common perturbation across different isotopes, similar to the QZS and the PLS. 
However, as discussed in our previous work~\cite{Ishiyama2026-wd}, the varying evaporation time required for each isotope can lead to small, differential temperature changes in the surrounding magnetic coils. 
Using the data of coil temperatures measured in Ref.~\cite{Ishiyama2026-wd}, we confirm that the residual BBR shift is well below the $10~\mathrm{mHz}$ level.

\textit{AOM chirp.}
The AOM chirp arises from a phase shift induced by switching the excitation laser on and off.
We adopt the same value $9(6)$~mHz as in Refs.~\cite{Ono2022-oy, Ishiyama2026-wd}, since the same product is employed for an AOM.

\section{Available Estimates of Selected SM Contributions}

\begin{table*}[!th]
    \renewcommand{\arraystretch}{1.5}
    \caption{Contributions of SM terms to the ISs $\dn{i}{174, 176}$ in units of MHz. From left to right, the columns list the values of the leading mass shift, leading field shift, next-leading-order Seltzer moment, quadratic field shift, and nuclear polarization $\epsilon_\text{pol}^{174,176}$. The values other than the nuclear polarization are derived from electronic factors based on atomic many-body calculations in Refs.~\cite{Kawasaki2024-co,Door2025-jv,Ishiyama2026-wd} and this work, as well as nuclear factors~\cite{Angeli2013-pc,Wang2021-ar,Door2025-jv}. The nuclear polarization contribution is taken from Ref.~\cite{Munro-Laylim2022-ho}. ``N/A'' indicates that the corresponding literature values are currently unavailable. See text for details.}
    
    \begin{ruledtabular}
    \label{tab: SM contribution}
    \begin{tabular}{lSSSSS}
        \multicolumn{1}{l}{Transition $i$} & \multicolumn{1}{c}{$K_i \w{174, 176}$} & \multicolumn{1}{c}{$F_i \dr{174, 176}$} & \multicolumn{1}{c}{$G_i^{(4)} \drr{174, 176}$} & \multicolumn{1}{c}{$G_i^{(2)} \ddr{174, 176}$} & \multicolumn{1}{c}{$\epsilon_\text{pol}^{174,176}$} \\ 
        [0.5ex]
        \colrule
        411 (Yb$^+$, $4f^{14}6s\;{}^2S_{1/2} \leftrightarrow 4f^{14}5d\;{}^2D_{5/2}$) & -2.0e1 & 2.0e3 & 6.0e1 & 6.5e-1 & 4e-3 \\ 
        431 (Yb, $4f^{14}6s^2\;{}^1S_0 \leftrightarrow 4f^{13}5d6s^2 \ (J=2)$) & 6.7e2 & -2.2e3 & {N/A} & -7.3e-1 & {N/A} \\ 
        467 (Yb$^+$, $4f^{14}6s\;{}^2S_{1/2} \leftrightarrow 4f^{13}6s^2\;{}^2F_{7/2}$) & 6.7e2 & -4.5e3 & -1.3e2 & -1.5 & {N/A} \\ 
        507 (Yb, $4f^{14}6s^2\;{}^1S_0 \leftrightarrow 4f^{14}6s6p\;{}^3P_2$) & {N/A} & 1.2e3 & {N/A} & 4.0e-1 & {N/A} \\ 
        578 (Yb, $4f^{14}6s^2\;{}^1S_0 \leftrightarrow 4f^{14}6s6p\;{}^3P_0$) & -3.5e1 & 1.2e3 & 4.9e1 & 4.0e-1 & {N/A} \\     
    \end{tabular}
    \end{ruledtabular}
\end{table*}

As discussed around Eq.~(1) in the main text, the ISs of optical transitions in Yb can be decomposed into several contributions within the framework of the SM.
Table~\ref{tab: SM contribution} summarizes theoretical estimation of the five SM terms for $\dn{i}{174, 176}$ of the five transitions used in this work.
The leading mass shift $K_i \w{174, 176}$ is rewritten as $K'_i m_{172} \w{174, 176}$, where the values of $K_i'$ and $\w{174, 176}$ are taken from Refs.~\cite{Kawasaki2024-co} and \cite{Door2025-jv}, respectively.
The nuclear mass of $^{172}$Yb, $m_{172}$, is derived from the atomic mass measurement in Ref.~\cite{Wang2021-ar}, accounting for the electron mass and binding energy as detailed in Ref.~\cite{Ono2022-oy}.
For the leading field shift $F_i \dr{174, 176}$ and the quadratic field shift $G_i^{(2)} \ddr{174, 176}$, the electronic factors are taken from Ref.~\cite{Ishiyama2026-wd} except for the 507-nm transition ($F_{507}$ and $G_{507}^{(2)}$), which are calculated in this work (see Sec.~\ref{sec:calc}). These factors are used in conjunction $\dr{174, 176}$ values from Ref.~\cite{Angeli2013-pc}, choosing $^{176}$Yb as the reference isotope $A_0$ for $\ddr{}$.
The next-leading-order Seltzer moment contributions $G_i^{(4)} \drr{174, 176}$ are calculated using the fiducial $G_i^{(4)}$ values in Ref.~\cite{Door2025-jv}.
For $\drr{174, 176}$, we adopt the weighted average of theoretical models `1.8/2.0 (EM), VS1' and `$\Delta \mathrm{N}^2 \mathrm{LO}_{\mathrm{GO}}$, VS1', as done in Ref.~\cite{Door2025-jv}.
Finally, for nuclear polarization, we quote the theoretical estimate for the 411-nm transition reported in Ref.~\cite{Munro-Laylim2022-ho}.
As shown in Table~\ref{tab: SM contribution}, the dominant sub-leading effects are $G_i^{(4)} \drr{174, 176}$ followed by $G_i^{(2)} \ddr{174, 176}$.
For the 411-nm transition, the nuclear polarization contribution is orders of magnitude smaller than the other four terms.
In this work, we assume that the nuclear polarization contributions for the other four transitions are of a similar order of magnitude.
To the best of our knowledge, theoretical estimates of nuclear polarization for transitions other than the 411-nm and 436-nm (${}^2S_{1/2} \leftrightarrow {}^2D_{3/2}$) transition in Yb$^+$ are currently unavailable, and evaluating these remains an important task for future studies.

In the King plot analyses, while understanding the absolute magnitudes of individual contributions provides a necessary baseline, the crucial factor is the magnitude of the induced nonlinearity.
In fact, Ref.~\cite{Munro-Laylim2022-ho} theoretically investigates the 2D King nonlinearity, showing that the nonlinearity induced by nuclear polarization is estimated to be smaller than those from $G_i^{(4)} \drr{}$ and $G_i^{(2)} \ddr{}$ for the 411-nm transition. 
To date, no theoretical calculations have been reported for the nonlinearity in the 3D generalized King plot, representing a key area for future theoretical investigations.

\section{Calculation and Uncertainty estimation of electronic factors}\label{sec:calc}
\subsection{Definition of the electronic factors}
The electronic factors for the field shift can be defined as follows~\cite{Counts2020-ca}:
\bea
F_{i,X} &=& \frac{Z e^2}{4 \pi \varepsilon_0} \frac{\xi_{i, X}^{(0)}}{6},  \\ 
G_{i,A_0}^{(2)} &=&\left.\frac{1}{2}\frac{\partial F_{i,X}}{\partial \braket{r^2}^{X}} \right|_{X=A_0}, 
\eea
where $Z$ is the charge of the atom, $e$ is the elementary charge, $\varepsilon_0$ is the permittivity of vacuum. $\xi_{i, X}^{(0)}$ is the electron density at the origin and $\braket{r^m}^{X}$ is the $m$th nuclear charge moment of reference isotope $X$ (In our calculation uses $X=174$). 

In the point-nuclear-charge approximation, the interaction potential between an electron and a neutron mediated by a new boson is expressed by 
\begin{equation}\label{eq:Yukawa}
V(r) = \alpha_\mr{NP} V_\mr{Y}(r); \quad V_\mr{Y}(r)=\hbar c\frac{e^{-m_{\phi}cr/\hbar}}{r}\,,
\end{equation}
where $r$ is the distance of an electron from the center of the nucleus, $\alpha_{\mathrm{NP}}$ is defined as $(-1)^{1+s_\phi} y_e y_n / (4 \pi \hbar c)$, and $s_\phi$ and $m_{\phi}$ are the spin and mass of the new boson, respectively.
The coupling constant between the new boson and an electron (neutron) is denoted by $y_{e(n)}$.
The corresponding electronic factor $X_i$ is defined by~\cite{Delaunay2017PRD,Mikami2017-oh}
\be
X_i = \braket{\Psi_\mr{u}|V_\mr{Y}|\Psi_\mr{u}} - \braket{\Psi_\mr{l}|V_\mr{Y}|\Psi_\mr{l}},
\ee
where the index $i$ refers to the transition between the upper electronic state $\ket{\Psi_\mr{u}}$ and the lower state $\ket{\Psi_\mr{l}}$. 
More details on the electronic factors were summarized in Sec. IV-A of Supplementary Information of Ref.~\cite{Ishiyama2026-wd}.

We employ the finite-field perturbation theory to obtain the expectation value of $V_\mr{Y}$ at the equation-of-motion coupled-cluster singles and doubles (EOM-CCSD) level~\cite{Pople1968JCP_FFPT,Norman2018_property,Kuroda2022PRA}. The electronic energy is calculated based on the Hamiltonian that consists of the unperturbed electronic Hamiltonian $\hat{H}_0$ and the target operator $\hat{O}$,
\be
\hat{H}(\lambda)=\hat{H}_0+\lambda \hat{O},
\ee
where $\lambda$ is the perturbation strength.
Using the associated energy with the above Hamiltonian ($E(\lambda)$), the expectation value of the property can be expressed as follows:
\be
\braket{\hat{O}}=\left.\frac{d E(\lambda)}{d \lambda}\right|_{\lambda=0}.
\ee
We use the six-point central difference formula~\cite{Fornberg1988MC_FFPT,Knecht2011TCA_Mossbauer} for the numerical differentiation
\be\label{eq:six_point}
\braket{\hat{O}} \sim \frac{-E^{-3}+9 E^{-2}-45 E^{-1}+45 E^{+1}-9 E^{+2}+E^{+3}}{60 \lambda^{\text {opt }}},
\ee
where $E^n$ refers to $E(n\lambda^\mr{opt})$. 
To obtain the difference between the electronic factors in the ground and excited states in a single step, the excitation energy obtained at the EOM-CCSD level is taken as $E^n$. The employed step sizes are as follows: $\lambda_\mathrm{opt} = 1.0\times10^{-5}$ for $1.0 \leq m_{\phi} \leq 1.0\times10^6$, $\lambda_\mathrm{opt} = 1.0\times10^{-3}$ for $1.0\times10^6 < m_{\phi} \leq 1.0\times10^7$, and $\lambda_\mathrm{opt} = 0.1$ for $1.0\times10^7 < m_{\phi} \leq 1.0\times10^8$, where $\lambda_\mathrm{opt}$ is in a.u. and $m_{\phi}$ is in eV.
The analytical integration modules are employed to obtain $\braket{V_\mr{Y}}$ at the general-active-space configuration interaction (GASCI)~\cite{Knecht_thesis} and average-of-configuration Hartree-Fock (AOC-HF) levels.

\subsection{Computational details}  
The DIRAC program package~(githash: d764921d6)~\cite{saue2020dirac,DIRAC22} is used for all electronic structure calculations otherwise explicitly described. We employ the three kinds of relativistic Hamiltonians: Dirac-Coulomb (DC), Dirac-Coulomb-Gaunt (DCG), or the exact two-component molecular-mean field (X2Cmmf) Hamiltonian~\cite{Sikkema_Visscher_Saue_Ilias_2009} based on the DC Hamiltonian ($^2$DC$^M$).
The two-electron integrals for the small components (SS$\mid$SS) were explicitly included.
Unless otherwise explicitly stated, the Gaussian nuclear charge model~\cite{visscher1997ADNDT} is employed for the electron–nucleus interaction potential.

%
%
For the calculations of $F_i$ and \Gtwo\ for the 507 nm transition, the electron-excitation EOM-CCSD (EE-EOM-CC) method is employed, otherwise explicitly noted. 
The reference states of the EOM-CC calculations at the HF and CCSD levels are calculated with the electron configuration of $4f^{14}6s^{2}$. We turn on the \texttt{.OVERLAP} option in the EOM-CC calculation to obtain the $^3P_2$ state stably. 
%
We employ the dyall.ae3z and dyall.ae4z basis sets, including the tight exponents listed in Table S2 of Ref.~\cite{Ishiyama2026-wd} (three $s$ functions and two $p$ functions, noted as ae3z+3s2p and ae4z+3s2p in later sections). 
The nuclear model effect ($\Delta$Fermi) was obtained at the (AOC-)HF level using the GRASP code \cite{dyall1989grasp}. The density at the initial grid point ($r\sim1.4\times10^{-8} \; a_0$, where $a_0$ refers to the Bohr radius) is used for estimating the electron density at the origin.
The electron configurations employed for the $^3P_2$ and $^1S_0$  states are $4f^{14}6s6p_{3/2}$ and $4f^{14}6s^2$, respectively.
The values of $\Delta$Fermi listed in Table \ref{tbl:F_G_507} are obtained from the relative corrections and the values of baseline.
The high-order relativistic effect ($\Delta$Gaunt) is obtained at the (AOC-)HF level using the same electronic configurations employed for the $\Delta$Fermi evaluation.
More details of the computational methodology employed are provided in Table \ref{tbl:comp_det_507}.

Our calculations of $X_i$ can be categorized into the baseline and corrections. 
Potentially important corrections for these states are identified, referring to the uncertainty estimation in our previous study~\cite{Ishiyama2026-wd}, and these corrections are then added to the baseline. For the basis-set correction and core-valence correlations, the $X_i$ calculations are performed for $m_\phi =10^n \; (n=0,1,\ldots,8)$ eV, and the uncalculated mass regions are interpolated using a linear function. If otherwise explicitly noted, we employ the dyall.v3z and dyall.v4z basis sets~\cite{Dyall_4f} with tight exponents obtained in Ref.~\cite{Ishiyama2026-wd} (v3z+3s2p, v4z+3s2p) for the calculations using the DIRAC program.

\textit{Baseline.}
For the $X_i$ calculations of the 578 and 507 nm transitions, we employ the EE-EOM-CCSD method~\cite{shee2018equation} with the reference electronic configuration of $4f^{14}6s^{2}$. For the 411 nm transitions, we employ the electron-attachment EOM-CCSD (EA-EOM-CCSD) method with the reference electronic configuration of $4f^{14}$. The EOM-CCSD calculations are performed based on the $^2$DC$^M$ Hamiltonian. The electrons in the orbitals higher than $3p$ are correlated, and the virtual orbitals are truncated at 100 $E_\mr{h}$.

The 431 and 467 nm transitions are calculated using the GASCI method~\cite{Fleig2003JCP,Knecht2010JCP}. The molecular orbitals for the \Szr, \JJ, \Stw, and \FF\ states are calculated at the (AOC-)HF level~\cite{Thyssen_thesis} with the electronic configurations of $4f^{14}6s^2$, $4f^{13}6s^25d^1$, $4f^{14}6s^1$, and $4f^{13}6s^2$, respectively. 
Our correlation model of the GASCI calculation is labeled as kk(1)\_mm(2)\, where one and two holes are allowed in the orbital shells indicated by kk and mm, respectively. For example, the 4f(1)\_5s5p6s(2) model indicates that one hole is allowed in $4f$ orbitals and two holes in total are allowed in the $5s5p6s$ orbital shell.
In addition, we impose one hole in the $4f$ orbitals for the \JJ\ and \FF\ states, and we add one electron in the $5d$ orbitals for the \JJ\ state. 
More details are described in Table S1 of Ref.~\cite{Ishiyama2026-wd}. For the baseline, the 4f(1)\_5s5p6s(5d)(2) model is employed, where (5d) refers to the occupation number in the $5d$ orbital being allowed only for the \JJ\ state, and $5d$ orbitals are included in the virtual orbitals in the other states. The virtual orbitals are truncated at 30 $E_\mr{h}$, and the DC Hamiltonian is employed.

\textit{Basis set.}
For the 578, 507, and 411 nm transitions, the EOM-CCSD calculations are carried out using the v4z+3s2p basis sets. The other computational details are the same as those employed in the baseline. The basis-set extrapolation is not performed because reliable extrapolation schemes for atomic properties in electronic excited states have not been reported.

\textit{Core-valence correlation.}
For the 431 nm transitions, the contribution from the core-valence correlation effect is obtained from the difference between the 4s5s5p4f(1)\_6s(5d)(2) and 5s5p4f(1)\_6s(5d)(2) models with the truncation of the virtual orbitals at 10 $E_\mr{h}$. 
For the 467 nm transitions, the core-valence correlation effect is obtained from the difference between the 3s4s5s5p4f(1)\_6s(5d)(2) and 5s5p4f(1)\_6s(5d)(2) models with the truncation of the virtual orbitals at 30 $E_\mr{h}$. 
The v3z+3s2p basis sets and DC Hamiltonian are employed for these calculations.

\textit{High-order relativistic effects.}
The uncertainties of the relativistic effects are estimated from the difference between the DCG and DC Hamiltonian at the (AOC-)HF level with the v3z+3s2p basis sets. The AOC-HF calculations are performed with the electron configurations of $4f^{14}6s^16p^1_{1/2}$, $4f^{14}6s^16p^1_{3/2}$, $4f^{13}6s^25d^1$, $4f^{14}6s^1$, $4f^{14}5d^1_{5/2}$, and $4f^{13}6s^2$ for the \PP, \Ptw, \JJ, \Stw, \DD, and \FF\ states, respectively. 
The \Szr\ state is calculated with the electronic configuration of $4f^{14}6s^2$ at the HF level.

\textit{Core electrons.}
The contributions from the core electrons not included in the correlation calculations are estimated at the (AOC-)HF level with the v3z+3s2p basis sets based on the DC Hamiltonian. The employed electronic configurations are the same as the ones described in the above paragraph. 
For example, in the case of the 507 nm transition, the contributions from $1s,2s,\ldots,3p$ orbitals are obtained from the difference between the (AOC-)HF calculations with the electronic configurations of $4f^{14}6s^2$ and $4f^{14}6s^16p^1_{3/2}$. 
For the 431 and 467 nm transitions, the contributions of the electrons that are not included in the core-valence correlations are added in this section. For example, in the case of 431 nm transition, the contributions from the $1s,2s,\ldots,3p,4p,4d$ orbitals are obtained at the difference between (AOC-)HF calculations with the electronic configurations of $4f^{14}6s^2$ and $4f^{13}6s^25d^1$. 

\textit{Nuclear charge distribution.}
The corrections of the nuclear charge-distribution model are evaluated from the difference between the calculations using the Fermi-type and Gaussian-type nuclear models at the AOC-HF level based on the DC Hamiltonian. The GRASP code \cite{dyall1989grasp} is employed for these calculations. The point-nuclear model is employed for the Yukawa-type potential, as shown in Eq.~(\ref{eq:Yukawa}). The electronic configurations used for the GRASP calculations are the same as those used in the above correction calculations at the (AOC-)HF level.

\subsection{Final values and the associated uncertainties} \label{sec: Final value and uncertainty}  
The details of each correction, final value, and the associated uncertainty of $F_{507}$ and $G^{(2)}_{507}$ are summarized in Tables~\ref{tbl:comp_det_507} and~\ref{tbl:F_G_507}. These tables are analogs of Tables S3-S6 of Ref.~\citenum{Ishiyama2026-wd}.

The final values of $X_i$ are obtained by summing the baseline and the corrections (high-order relativistic effects, core electrons, and nuclear charge distributions). The contributions from the larger basis sets and the core-valence correlations are also added for the EOM-CCSD and GASCI calculations, respectively. The uncertainties are obtained from the Euclidean
norm of the individual uncertainties, except for the core-electron parts. 
Assuming that the electron correlation effects are small and the orbital-relaxation effects are dominant in the core-electron contributions, we incorporate the core-electron contributions into the scheme of Euclidean norm after halving these contributions. 
The relatively large contributions of relaxation effects in the core orbitals have been reported in calculations of core-electron ionization (e.g., \cite{Bellafont2015JCP_Delta_Koopmans,South2016PCCP,Hirao2025JPCA_core}).

\begin{figure}[!t]
    \centering
    \includegraphics[width=0.95\linewidth]{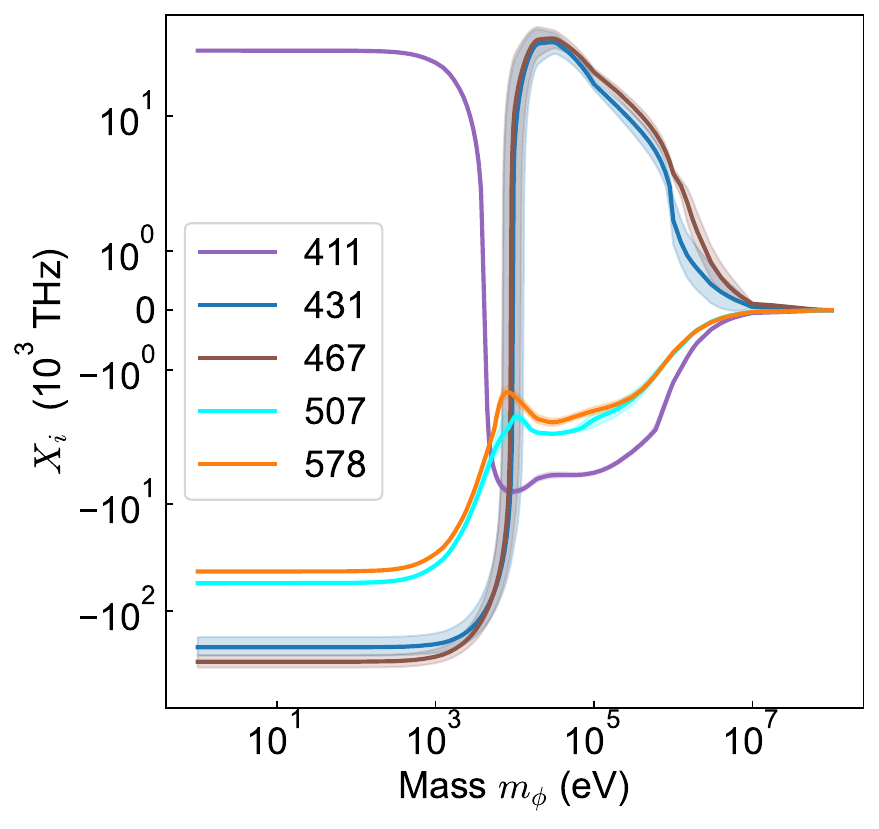}
    \caption{PS electronic factor $X_i$ calculated in this study as a function of the new-particle mass $m_{\phi}$. The solid lines show the final values of $X_i$ and the shaded area indicates the associated uncertainties.}
    \label{fig:X}
\end{figure}

In our uncertainty estimation scheme, the relative error becomes significant when cancellation occurs between the baseline (valence electrons) and core electrons. 
This large relative uncertainty is especially observed in the 431 and 467 nm transitions around 10 keV. The sign of $X_i$ of the baseline is changed between 10 and 20 keV, that is, the absolute value of $X_i$ of these transitions becomes very small around these mass regions (cf. kink structure of the $X_i$ curve visualized in Fig.~\ref{fig:X}). Meanwhile, the sign of the core-electron contributions is positive in all mass regions. This cancellation between the baseline (negative) and core-electron contribution (positive) results in a large relative uncertainty at around $m_{\phi} \sim 10$ keV. This large uncertainty might be the reason why the possibility of the new particle cannot be excluded around the mass of 9.67 keV.

%
\begin{table*}[!htbp]
\caption{
Summary of $F_i$ ($\mathrm{GHz} / \mathrm{fm}^2$) and \Gtwo\ ($\mathrm{MHz} / \mathrm{fm}^4$) for the $i=507$ ($^3P_2$) transition obtained at various calculation levels, which are used for obtaining the corrections in Table \ref{tbl:F_G_507}. 
The virtual cutoff energy (virtual) is in $E_\mr{h}$.
The calculation conditions for the blank rows are the same as those for the rows immediately above them.
The correction terms written on only one line were obtained taking the difference from the baseline.
}\label{tbl:comp_det_507}
\begin{tabular}{ lccccc rc } 
\hline\hline
&Hamiltonian & basis & method & virtual & correlated elec. & $F_{507}$ & $G^{(2)}_{507}$\tabularnewline
\hline
baseline & DC & ae3z+3s2p & EOM-CC & 100 & all & --10.730 & 29.935\tabularnewline \rule[0mm]{0mm}{5mm}
$\Delta$Basis & DC & ae4z+3s2p & EOM-CC & 100 & all & --10.601 & 29.396\tabularnewline
 &  & ae3z+3s2p &  &  &  &  & \tabularnewline \rule[0mm]{0mm}{5mm}
$\Delta$Triple & DCG & ae3z+3s2p & CCSD(T) & 100 & higher than 3p & --15.640 & 43.269\tabularnewline
 &  &  & EOM-CC &  &  & --15.789 & 43.681\tabularnewline \rule[0mm]{0mm}{5mm}
$\Delta$Virt. cutoff & DC & ae3z+3s2p & EOM-CC & 3000 & all & --10.728 & 30.043\tabularnewline  \rule[0mm]{0mm}{5mm}
$\Delta$Gaunt & DCG & ae4z+3s2p & HF &  &  & --8.860 & 24.657\tabularnewline 
 & DC &  &  &  &  & --8.908 & 24.791\tabularnewline 
\hline\hline
\end{tabular}
\end{table*}

%
\begin{table}[!htbp]
\caption{
Calculated values of $F_i$ ($\mathrm{GHz} / \mathrm{fm}^2$) and \Gtwo\ ($\mathrm{MHz} / \mathrm{fm}^4$) for the 507 transition including the high-order corrections.
The corrections are obtained from Table \ref{tbl:comp_det_507}.
}\label{tbl:F_G_507}
\begin{tabular}{l SS}
\hline\hline
 & {$F_{507}$} & {$G^{(2)}_{507}$}\tabularnewline
 \hline
baseline & -10.730 & 29.935\tabularnewline
$\Delta$Basis & 0.129 & -0.539\tabularnewline
$\Delta$Triple & 0.102 & -0.285\tabularnewline
$\Delta$Virt. cutoff & 0.002 & 0.108\tabularnewline
$\Delta$Gaunt & 0.048 & -0.134\tabularnewline
$\Delta$Fermi & 0.173 & 0.392\tabularnewline
Final result & -10.277 & 29.478\tabularnewline
uncertainty & 0.243 & 0.745\tabularnewline
\hline\hline
\end{tabular}
\end{table}

\section{Statistical treatment of King plot analysis}
\subsection{Statistical test of the 3D generalized King linearity}
We employ a similar method to our previous works~\cite{Ono2022-oy, Ishiyama2026-wd} as the procedure of the statistical test.
To test the 3D King linearity with Eq.~(2) in the main text for a specific transition set, $\chi^2$ is defined as 
\begin{equation} \label{eq: chi2 3D individual}
    \chi^2 = \chi^2_{\eta} + \sum_{i=1}^{3} \chi^2_{\delta \nu_i}.
\end{equation}
Here, $\chi^2_{\eta}$ corresponds to nuclear mass ratios $\eta^A = m_A / m_{172}$~\cite{Door2025-jv}:
\begin{equation} \label{eq: chi2 mass}
    \chi^2_{\eta} = \sum_{A=168, 170, 174, 176} 
    \left\{ \frac{1 / \left( \w{A, A_0} - \w{172, A_0} + 1 \right) - \hat{\eta}^{A}}{\sigma_{\eta^{A}}} \right\}^2,
\end{equation}
where $\hat{(\cdot)}$ and $\sigma_{(\cdot)}$ denote the measured values and their experimental uncertainties, respectively.
The parameters $w^{A, A_0}$ are fitting parameters, where $A_0$ is the mass number of the reference isotope ${}^{176}$Yb.
The second term in Eq.~(\ref{eq: chi2 3D individual}) is corresponding to the ISs:
\begin{equation} \label{eq: chi2 IS}
    \chi^2_{\delta \nu_i} = \sum_{(A',A)} 
    \left\{ \frac{ \left( \dn{i}{A'A_0} - \dn{i}{AA_0} \right) - \hat{\delta \nu}_{i}^{A'A}}{\sigma_{\dn{i}{A'A}}} \right\}^2. 
\end{equation}
The sum is taken over the isotope pairs for which the ISs are actually measured.

Basically, there are 16 experimental values, since each of $\hat{\eta}^A$, $\hat{\delta \nu}_{1}^{A'A}$, $\hat{\delta \nu}_{2}^{A'A}$, and $\hat{\delta \nu}_{3}^{A'A}$ has four elements associated with four isotope pairs.
On the other hand, we define 15 fitting parameters composed of three King linearity parameters $k$, $f_1$, and $f_2$, as well as the other 12 parameters $\w{AA_0}$, $\dn{1}{AA_0}$, and $\dn{2}{AA_0}$.
Thus, the dof of the fit is $16 - 15 = 1$
~\footnote{For the 411 and 467-nm transitions, the directly measured quantities are the absolute transition frequencies of the five bosonic isotopes~\cite {Door2025-jv}, not the ISs of the four isotope pairs.
To account for this, we introduce one additional fitting parameter for the absolute frequency of the reference isotope $^{176}$Yb.
The $\chi ^2$ function is then adjusted accordingly to properly incorporate the statistical correlations among the resulting IS values.
Note that this treatment does not change the dof of the fit.}.
Note that the ISs of the third transition are not defined as independent parameters like $\dn{1}{AA_0}$ and $\dn{2}{AA_0}$, but are obtained by Eq.~(2) in the main text.

The aforementioned treatment is straightforwardly extended to a simultaneous fit of linearly independent three transition sets among the five transitions.
The fitting function is explicitly given as the following simultaneous equations:
\begin{equation} \label{eq: chi2 3D simultaneous}
\left\{ \,
    \begin{aligned}
    & \Dn{3} = k^{(3)} \bm{w} + f_1^{(3)} \Dn{1} + f_2^{(3)} \Dn{2} , \\
    & \Dn{4} = k^{(4)} \bm{w} + f_1^{(4)} \Dn{1} + f_2^{(4)} \Dn{2} , \\
    & \Dn{5} = k^{(5)} \bm{w} + f_1^{(5)} \Dn{1} + f_2^{(5)} \Dn{2} ,
    \end{aligned}
\right.
\end{equation}
where the superscript $(i)$ means that the respective fitting parameter corresponds to the King linearity of the transition set $(1, \ 2, \ i)$.
In a geometric sense, this simultaneous analysis is equivalent to fitting the four isotope-pair data points within the 5D transition space with a single 2D subspace (i.e., a plane).
Here, the 21 fitting parameters are defined, and the dof increases to 3.
The definition of $\chi^2$ is then modified as
\begin{equation}
    \chi^2 = \chi^2_{\eta} + \sum_{i=1}^{5} \chi^2_{\delta \nu_i}.
\end{equation}

\subsection{Analysis for the new particle}
Next, we present how to obtain a favored region in the ($m_\phi$, $y_e y_n$) space for a specific transition set under the assumption that the IS is composed of the leading-order mass shift, the leading-order field shift, one factorizable higher-order SM term, and the PS.
The fitting function is changed from Eq.~(2) to Eq.~(3) in the main text, while the definition of $\chi^2$ is the same as Eq.~(\ref{eq: chi2 3D individual}).
In this analysis, $X_i$ is treated as a constant value based on the theoretical calculation, not a fitting parameter.
Since we define $\alpha_\mathrm{NP}$ as an additional fitting parameter, the dof decreases from 1 to 0, which prevents us from testing the underlying assumption in this fit model.
The obtained best-fit value and 95\% confidence interval of $\alpha_\mathrm{NP}$ are displayed as the solid line and shaded area in Fig.~3(a) in the main text.

For the simultaneous fit, Eq.~(3) in the main text is spanned over the three transition sets in a manner similar to the previous section.
The fitting function is explicitly expressed as
\begin{equation} \label{eq: 3D PS simultaneous}
\left\{ \,
    \begin{aligned}
    \Dn{3} = & k^{(3)} \bm{w} + f_1^{(3)} \Dn{1} + f_2^{(3)} \Dn{2} \\
             & + \alpha_{\mathrm{NP}} (X_3 - f_1^{(3)} X_1 - f_2^{(3)} X_2) \DA, \\
    \Dn{4} = & k^{(4)} \bm{w} + f_1^{(4)} \Dn{1} + f_2^{(4)} \Dn{2} \\
             & + \alpha_{\mathrm{NP}} (X_4 - f_1^{(4)} X_1 - f_2^{(4)} X_2) \DA , \\
    \Dn{5} = & k^{(5)} \bm{w} + f_1^{(5)} \Dn{1} + f_2^{(5)} \Dn{2} \\
             & + \alpha_{\mathrm{NP}} (X_5 - f_1^{(5)} X_1 - f_2^{(5)} X_2) \DA .
    \end{aligned}
\right.
\end{equation}
Notably, since the fitting parameter $\alpha_{\mathrm{NP}}$ is common for the three transition sets, the fit retains 2 dof. 
This allows us to test the hypothesis that the IS consists of the three SM terms and the PS, as shown by the blue solid line in Fig.~3(b) in the main text.

To account for the $X_i$ uncertainties, all five $X_i$ in Eq.~(\ref{eq: 3D PS simultaneous}) are replaced with fitting parameters, instead of being treated as fixed constants.
Then, the corresponding $\chi^2$ term is additionally introduced as 
\begin{equation}
    \chi^2 = \chi^2_{\eta} + \sum_{i=1}^{5} \chi^2_{\delta \nu_i} + \sum_{i=1}^{5} \chi^2_{X_i}, 
\end{equation}
where 
\begin{equation}
    \chi^2_{X_i} = \left( \frac{X_i - \hat{X_i}}{\sigma_{X_i}} \right)^2.
\end{equation}
$X_i$ is the additionally defined fitting parameter, while $\hat{X_i}$ and $\sigma_{X_i}$ are the theoretical best value and its $1\sigma$ uncertainty, respectively.
Note that $\hat{X_i}$ and $\sigma_{X_i}$ are uniquely determined once the mass $m_\phi$ is specified.
Then, the resultant $\chi^2$ values are plotted as a function of $m_\phi$ in Fig.~3(b) in the main text, yielding the minimum value 0.04 at $m_\phi = 9.67$~keV.

Finally, regarding $m_\phi$ as yet another fitting parameter, the contour plot of the 95\% confidence level is obtained in the ($m_\phi$, $y_e y_n$) space, as shown by the red solid line and shaded area in Fig.~3(a) in the main text.
This region corresponds to the area below the $\chi^2$ value at the 95\% confidence level in Fig.~3(b) in the main text.

\clearpage
\bibliography{reference}